\documentclass[12pt]{spieman}  
\usepackage{amsmath,amsfonts,amssymb}
\usepackage{graphicx}
\usepackage{setspace}
\usepackage{tocloft}
\usepackage{tabularx}
\usepackage{array}
\usepackage{multirow}
\usepackage{lineno}
\usepackage{xcolor}

\def\change#1{{#1}}

\title{Development of a planar cable-driven parallel robot for submillimeter and terahertz beam mapping measurements}

\author[a,*]{Evan~C.~Mayer}
\author[a]{Ian~N.~Lowe}
\author[a]{Daniel~P.~Marrone}

\author[c,e]{James~J.~Bock}
\author[c,e]{Charles~M.~Bradford}
\author[d]{Victoria~L.~Butler}
\author[e]{Tzu-Ching~Chang}
\author[c,e]{Yun-Ting~Cheng}
\author[d]{Dongwoo~T.~Chung}
\author[d,c]{Abigail~T.~Crites}
\author[h]{Audrey~Dunn}
\author[a]{Nicholas~Emerson}
\author[e]{Clifford~Frez}
\author[e]{Jonathon~Hunacek}
\author[f]{Ryan~P.~Keenan}
\author[b]{Chao-Te~Li}
\author[c]{King~Lau\footnote{deceased}}
\author[g]{Guochao~Sun}
\author[a]{Isaac~Trumper}
\author[e]{Anthony~D.~Turner}
\author[d]{Benjamin~Vaughan}
\author[b]{Ta-Shun~Wei}
\author[h]{Michael~Zemcov}

\affil[a]{Department of Astronomy and Steward Observatory, University of Arizona, 933 N Cherry Avenue, Tucson, AZ 85721, USA}
\affil[b]{Institute of Astronomy and Astrophysics, Academia Sinica, Taipei, Taiwan}
\affil[c]{Department of Physics, California Institute of Technology, Pasadena, California 91125, USA}
\affil[d]{Department of Physics, Cornell University, Ithaca, NY, 14853, USA}
\affil[e]{Jet Propulsion Laboratory, California Institute of Technology, Pasadena, California 91109, USA}
\affil[f]{Max-Planck-Institut für Astronomie, Königstuhl 17, D-69117 Heidelberg, Germany}
\affil[g]{CIERA and Department of Physics and Astronomy, Northwestern University, 1800 Sherman Avenue, Evanston, IL 60201, USA}
\affil[h]{Rochester Institute of Technology, Rochester, NY, 14623, USA}

\cftpagenumbersoff{figure}
\cftpagenumbersoff{table} 
\begin{document} 
\maketitle

\begin{abstract}

The spatial sensitivity pattern of millimeter-wavelength receivers is an important diagnostic of performance and is affected by the alignment of coupling optics. Characterization can be challenging in the field, particularly in the decentered and tightly packed optical configurations that are employed for many astronomical millimeter-wave cameras.
In this paper, we present the design and performance of a lightweight and reconfigurable beam mapper, consisting of a bank of thermal sources positioned by a planar cable-driven robot. We describe how the measurement requirements and mechanical constraints of the Tomographic Ionized-carbon Mapping Experiment (TIME) optical relay drive the design of the mapper. To quantify the positioning performance, we predict the beam patterns at each surface to derive requirements and use a non-contact computer-vision based method built on OpenCV to track the payload position with an accuracy better than 1.0~mm. We achieve an in-plane absolute payload position error of 2.7~mm (RMSE) over a $\sim$400~mm $\times$ 400~mm workspace and an in-plane repeatability of 0.81~mm, offering substantial improvements in accuracy and speed over traditional handheld techniques.

\end{abstract}


{\noindent \footnotesize\textbf{*}Evan C. Mayer,  \linkable{evanmayer@arizona.edu} }


\section{Introduction} \label{sec:intro}

The verification and optimization of alignment and focus is a critical task in the commissioning of an optical system. At visible wavelengths, many tools are available to generate and capture light at various positions in the system. The necessarily high precision of the optical elements and their large size relative to the wavelength of interest also enables the use of lasers to directly trace rays through the system. Millimeter and submillimeter imaging systems use highly specialized detectors and tend to operate with optical elements that are smaller (in wavelength units) and therefore diffract more, and contribute thermal glow that overwhelms many sources of illumination. As a consequence, very different approaches are generally required to validate the optical alignment and performance.

One approach that can be used with millimeter-wave systems is the direct measurement of mirror positions. Although the mirror surfaces themselves may be hard to precisely characterize, reference features such as retroreflector and tooling ball nests on the mirror perimeters can provide an initial relative alignment. Capturing the optical propagation path requires further work, however. Previous experiments have approached this in various ways, such as replacing some mirrors with partially optically reflective ones,  roughly tracing detector response with hot absorptive materials (LArge APEX BOlometer CAmera\cite{siringo_large_2009}), sighting with alignment scopes, hot-wire crosshairs, and near-infrared cameras (Submillimeter Common-User Bolometer Array 2 Fourier Transform Spectrometer 2\cite{gom_testing_2010}), or the lowest-effort and least repeatable method, moving a radiative or absorptive source across the path of the receiver's beam by hand. Coherent receivers often use a monochromatic transmitter on a movable stage\cite{tong_near_2003, kim_tilted_2018, davis_complex_2019} to perform vector beam mapping, though the additional precision required often makes phase-sensitive measurements impractical in the field.

\subsection{Case Study: The Tomographic Ionized-carbon Mapping Experiment (TIME)} \label{sec:time}

\begin{figure}
    \centering
    \includegraphics[width=\textwidth]{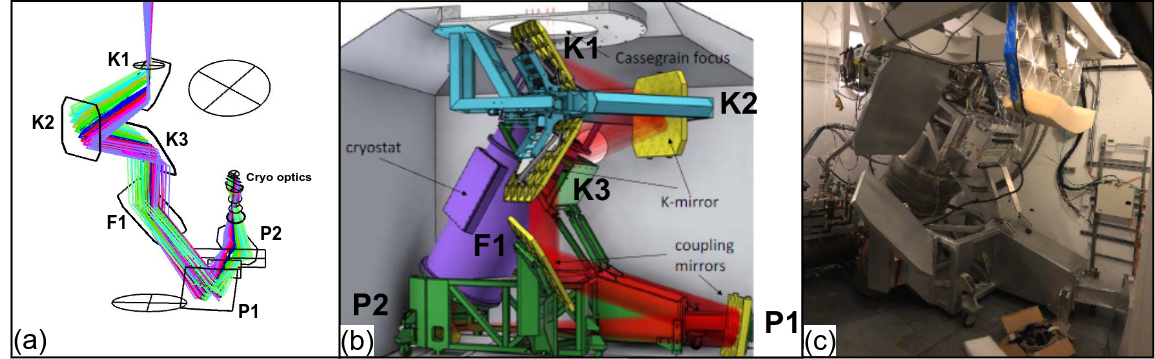}
    \caption{\label{fig:warm_optics} The TIME optical relay system for the Kitt Peak 12~m radio telescope. Light from the Cassegrain telescope enters from the top. It passes through a field derotator (K-mirror) comprising mirrors K1-3, before encountering the flat fold mirror F1, and two powered mirrors P1 and P2 near the cabin floor. The latter directs light upward to the cryostat window. (a) A Zemax OpticStudio ray trace of several feeds with the K-mirror displaced by 45$^{\circ}$ from the straight-ahead home position. (b) A mechanical rendering showing the notional path (red) of a beam through the optics, with the K-mirror rotated the opposite direction. Mirrors are highlighted in yellow. (c) The as-built optical system, installed in the ARO 12~m receiver cabin on Kitt Peak.}
\end{figure}

The Tomographic Ionized-carbon Mapping Experiment (TIME)\cite{crites_time-pilot_2014} is a millimeter-wave instrument with a complex optical system coupling it to the telescope. The primary science mission is line intensity mapping \cite{kovetz_line-intensity_2017, bernal_line-intensity_2022} (LIM) of [CII] fine structure cooling lines at redshifts $z\approx5-9$, and CO rotational transition lines at redshifts $z\approx0.5-2$.\cite{sun_probing_2021} The TIME receiver scans a line of feedhorns across the sky, illuminating a cryogenically cooled array of 1920 transition-edge sensor (TES) bolometers that operate at 186-324~GHz (1.61-0.93~mm). When deployed on the Arizona Radio Observatory (ARO) 12~m radio telescope at Kitt Peak, the telescope optics are coupled to the TIME cryostat's cold optics by a room-temperature system of relay mirrors (the ``warm optics"). This relay occupies nearly the entire volume of the receiver cabin (Fig.~\ref{fig:warm_optics}). It includes a motorized parallactic angle derotator\cite{trumper_utilizing_2019} (``K-mirror") to maintain the angle of the line of feedhorn beams on-sky, as well as flat and powered mirrors to redirect and efficiently couple the beams into the cryostat window. The optimization of the TIME optical system, which contains several free-form elements, is described in Ref.~\citenum{trumper_freeform_2019}. Light from the warm optics must enter the cryostat window at the proper place and angle to propagate through a polyethylene lens, filter stack, wire grid polarizer, spatial array of feedhorns, and a grating spectrometer before illuminating the detector arrays.\cite{hunacek_design_2016, hunacek_detector_2016, li_time_2018}

TIME is installed on the telescope for only part of the year, and alignment of the optical system must be re-verified each observing season. The warm optics are aligned to each other with a laser tracker, but without additional equipment, their alignment to the internal cryostat optics can only be verified through on-sky measurements.

The system described in this paper was designed to provide an automated and repeatable tool for measuring the profiles of the TIME beams as they are relayed through the optical system. The TIME optical configuration requires that the mapping device operate on mirrors of various sizes and shapes, with limited clearance, and in many orientations relative to the gravity vector. The beam sizes also vary significantly as a function of frequency and optical element. To overcome these difficulties, we have implemented a lightweight reconfigurable scanning mapper in the form of a planar cable-driven parallel robot (CDPR). It carries and controls an array of Hawkeye Technologies IR-50 infrared sources, which are chopped to make measuring their amplitude easier in noisy detector timestreams.

In this paper, we briefly outline the theoretical work on the statics problem that forms the basis for the mechanical design and numerical control algorithm. Then, we present the control algorithm, design parameters, and construction in Sec.~\ref{sec:desc}. We conduct simulations to derive requirements and perform experiments to evaluate the mapper's absolute position accuracy and repeatability in Sec.~\ref{sec:perf}. We evaluate sources of error and discuss the mapper's suitability to the task in Sec.~\ref{sec:disc}.

\section{Instrument Description} \label{sec:desc}

The primary requirements driving the mapper design were the ability to reference mapper positions to mirror locations and achieve positioning accuracy sufficient to sample the beam patterns. A secondary goal was to create a design that obscured the minimum amount of each mirror, in order to limit parasitic thermal emission and increase contrast with the chopped infrared sources. To achieve these goals, we took inspiration from the world of parallel manipulators\cite{gough_universal_1962, stewart_platform_1965, cappel_motion_1967}, and designed a cable-driven parallel robot (CDPR). CDPRs excel at moving payloads precisely over large workspaces with a minimum of support structure. For example, Skycam\cite{brown_suspension_1987} rapidly moves a broadcast television camera over live sporting events to give better coverage, the National Institute of Standards and Technology (NIST) Robocrane\cite{albus_nist_1992} was envisioned for shipbuilding and pipe fitting, and Five-hundred-meter Aperture Spherical radio Telescope (FAST)\cite{fast_collaboration_commissioning_2019} is an enormous radio telescope with a suspended receiver platform which is steerable to observe different parts of the sky. Projects with similar requirements have shown the feasibility of a static frame design with actuators at the corners\cite{ottaviano_low-cost_2005, jin_four-cable-driven_2013,gosselin_initial_2018, jomartov_development_2021} to achieve fast and precise planar motion.

\begin{figure}
    \centering
    \includegraphics[width=\textwidth]{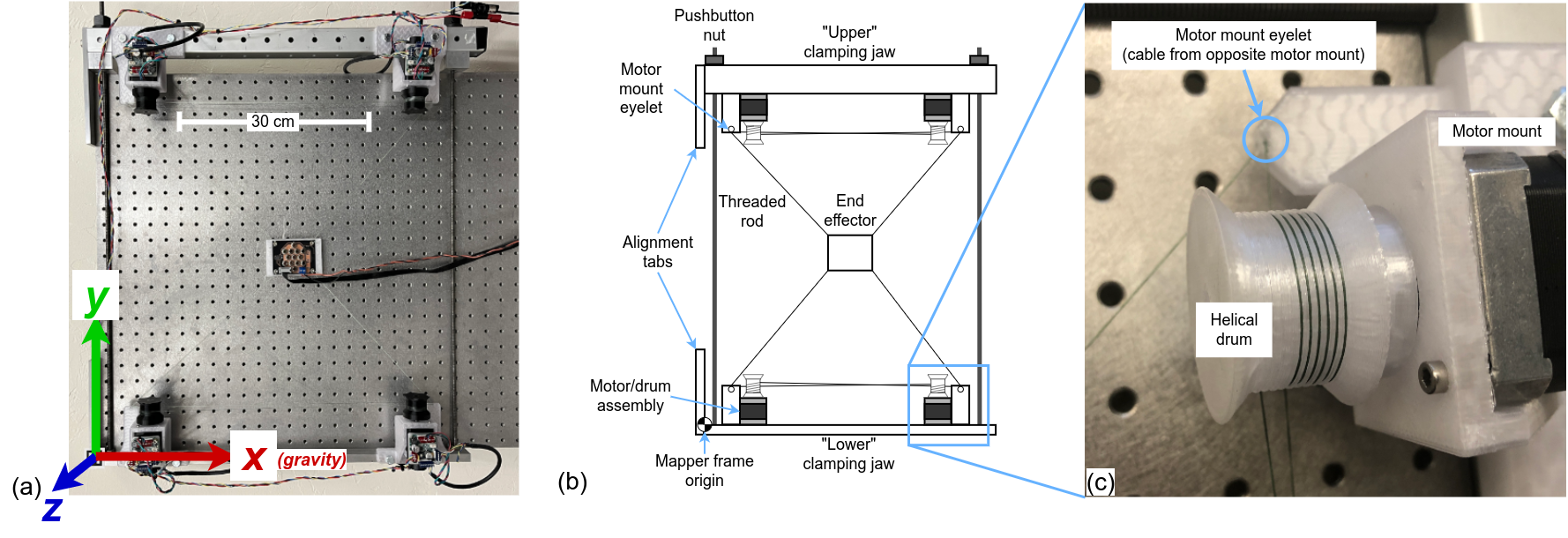}
    \caption{\label{fig:full_system} (a) The mirror mapper, clamped on a 600~mm wide optical breadboard. The gravity vector points along +$x$ in this setup. For scale, holes are located on a 1-inch (25.4~mm) grid pattern. At the top and bottom, aluminum extrusions provide mounting points for four stepper motors with encoders. Cable for each axis is wound onto helical drums directly driven by the stepper motors. Eyelets that control cable routing are integrated with the motor mounts. The end effector, a PCB raft of modulated infrared emitters, is suspended by the fine cables in the center of the workspace. The black and red wires hanging to the right are power and control wires for the infrared emitters. The blue, yellow, black and red twisted wires are the RS-485 bus carrying commands from the control computer to the motor driver boards. (b) A labeled schematic diagram of the system in (a). (c) A close-up view of the helical drum and motor mount with integrated cable eyelet.}
\end{figure}

A breakdown of the mapper system is presented in Fig.~\ref{fig:full_system}. We describe the frame and support structure in Sec.~\ref{sec:frame}, the actuation system in Sec.~\ref{sec:drive}, the end effector in Sec.~\ref{sec:payload}, the numerical control algorithm in Sec.~\ref{sec:control}, and homing in Sec.~\ref{sec:homing}.

\subsection{Frame and Support Structure} \label{sec:frame}

The physical layout of the mapper is determined by a number of factors: space constraints, usable workspace, payload attitude, payload capacity, and absolute accuracy. The relevant properties of the warm mirrors are summarized in Table \ref{tab:mirrors}. The mirror envelopes are rectangular in shape, with widths ranging from 470-800~mm, and heights ranging from 430-650~mm. The lack of space underneath the TIME cryostat fully restricts access to P2, so we cannot map it, and the final range of shapes to accommodate is 600-800~mm and 460-650~mm. Accessibility of reference surfaces is essential to translate mapper movements and receiver timeseries data into beam maps in terms of position across each mirror.  The proximity of mirrors K1-K3 to the K-mirror support structure restricts access to several edges that could be used as reference surfaces, and all mirrors except P1 have corners machined off (see Fig.~\ref{fig:warm_optics}), leaving only edges as reference features.

\begin{table}
\begin{tabularx}{\textwidth}{l|X|X|l}
    \textbf{Mirror} & \textbf{Envelope Width \newline (mm)} & \textbf{Envelope Height \newline (mm)} & \textbf{Comments} \\
    \hline
    K1 & 600 & 460 & High in cabin poor lower edge access \\
    K2 & 650 & 650 & Poor lateral edge access \\
    K3 & 750 & 550 & Angled $\sim45$~deg toward floor, poor upper edge access \\
    F1 & 800 & 460 & Angled toward ceiling \\
    P1 & 600 & 500 & Poor edge access near floor \\
    P2 & 470 & 430 & No access \\
    \hline
\end{tabularx}
\caption{\label{tab:mirrors} A table of mirrors in the optical system, with their sizes and relevant physical quantities, as well as comments on areas with access restrictions. The size of the rectangular envelope is given as the distance between pairs of parallel sides. These properties determine the design of the mapper's clamping frame: it must open wide enough to clamp on across the shortest accessible dimension of the largest mirror, it must not collide with the optical support structures or the receiver cabin walls or floor, and it should cover the largest workspace possible subject to these constraints.}
\end{table}

To accommodate the range of mirror sizes and maximize workspace over the surface of each mirror, we use an adjustable clamping design. The mapper support frame, constructed of extruded aluminum profile, is referenced to the mirror surface coordinate system by physically clamping to two opposing edges of a mirror. Clamping force is achieved by a pair of threaded rods with protective covering to prevent marring the mirror surface, spanning one dimension of the mirror and affixed to each side. The referencing of the mapper planar coordinate system to the mirror being mapped is completed by the addition of a pair of aluminum alignment tabs (Fig.~\ref{fig:full_system}a, top; Fig.~\ref{fig:full_system}b, left) to each aluminum profile that contact one perpendicular mirror edge. The upper and lower extruded aluminum profiles (Fig.~\ref{fig:full_system}b, top and bottom) provide support for four actuators and associated cable harnessing.

\subsection{Drive System} \label{sec:drive}

The drive system of any parallel manipulator consists of two or more actuators working together. The number of actuators needed for a CDPR is determined by the number of degrees of freedom needed for motion: two cables can move a platform along a line, three can constrain motion to a plane, and so on. Although only three are required to constrain motion in the scanned plane, we use four actuators. The extra actuator allows a greater workspace to be spanned \cite{chablat_comparative_2011} and provides an additional degree of freedom to control the in-plane rotation of the payload. At each corner, a cable actuator is composed of: (1) a Trinamic Motion Control GmbH QSH4218-10000-AT 2-phase stepper motor with 10000 line (40000 tick) resolution incremental encoder, (2) an AllMotion EZHR17EN Stepper Motor Control \& Driver board powered by 24 VDC, (3) a fused deposition modeling (FDM 3D printed)-manufactured motor mount with an integrated cable eyelet, printed in heat- and wear-resistant polyethylene terephthalate glycol (PETG) filament, and (4) a PETG drum with a helical groove to accommodate excess cable, fixed directly to the motor shaft. Stepper motors are operated via an RS-485 serial bus in closed-loop mode with 256x microstepping and encoder feedback.
Cables are fixed to each drum, and lengths are controlled for each axis by winding cable on or off of the helical drums. Controlled winding on and off of the helical drums is accomplished, contrary to prior art, not with a synchronized cable winding guide\cite{jin_four-cable-driven_2013} but simply by routing each cable through a distant eyelet located on an opposing motor's mount (see Fig.~\ref{fig:full_system}b and c). This ensures the cable always meets the drum at an angle nearly perpendicular to the axis of rotation, and consistent winding is achieved when cable tension is maintained. Choosing a sufficiently distant eyelet from the drum also lessens the effect of error incurred by the changing position of the cable on the drum as a function of motor angle, an approximation we explore in detail in Sec.~\ref{sec:errsrcs}.

The cable itself is braided ultra high molecular weight polyethylene fiber (also known by trade name Dyneema\textregistered), selected for wear resistance, stretch resistance, and availability for future maintenance. The type used is commercially available as SpiderWire EZ Braid$^{\textrm{TM}}$ Superline, with 22.6~kg breaking (pull test) strength rated by the manufacturer. Each cable is routed from the helical drum through a small eyelet located on a standoff integrated with the motor mount. Eyelets are offset $\sim$80~mm from the clamping jaws toward the interior of the workspace, and separated laterally by $\sim$500~mm. Each eyelet is a cylindrical hole in the printed PETG standoff, sized to closely match the diameter of the cable. This is critical, as it sets the corner anchor location for the geometry of the mapper cable system; sizing the eyelet to match the cable diameter as closely as possible without adding excess friction ensures that as the raft changes position in the workspace, the effective location of the corner remains constant. See a discussion of this type of error, which is typically handled by a careful arrangement of pivoting pulleys, in Refs. \citenum{landsberger_design_1984, kino_error_2018}.

\subsection{End Effector Design} \label{sec:payload}

\begin{figure}
    \centering
    \includegraphics[width=0.4\textwidth]{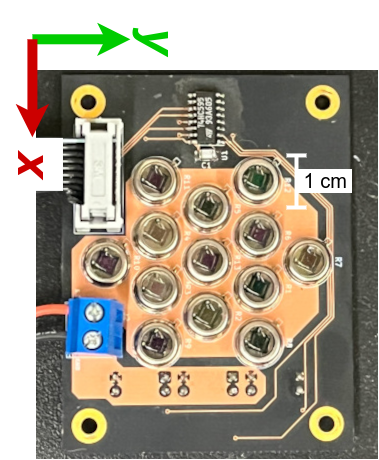}
    \caption{\label{fig:hawkeye_board} The end effector, consisting of 13 Hawkeye Technologies IR-50 pulsable emitters. The emitters are arranged in three groups: 1 center, 6 inner, and 6 outer. Each group is powered by 6.7~VDC from the blue screw terminal, and modulated by MOSFETs on the back of the board that are controlled via a Serial Peripheral Interface to the Texas Instruments SN74HC595 shift register. The radius of the outer ring of emitters is $\sim$18.5~mm.}
\end{figure}

The design of the end effector, or payload, is tied to considerations of the load bearing capacity of the mapper, placing constraints on the actuator power supply and motors and factoring in to the robustness of positioning the mapper frame relative to each mirror. The payload experiences external forces from gravity and moments from any power or communications cables that are not strain relieved. For this reason, we minimize the weight of the payload and the number and rigidity of the wires. The main body of the payload is composed of a custom printed circuit board (PCB) carrying three concentric rings of a total of 13 Hawkeye Technologies IR-50 pulsable infrared emitters in TO-5 packages (Fig.~\ref{fig:hawkeye_board}). Similar emitters have been used for detector identification on the BLAST-TNG balloon-borne telescope \cite{lowe_characterization_2020}, mapping detector responses for TolTEC \cite{wilson_toltec_2020}, and are being tested for in-flight calibration for the Terahertz Intensity Mapper \cite{marrone_terahertz_2022}. Upon reaching each commanded position, the IR-50s are blinked ten times at 5 Hz with a 50\% duty cycle, which is empirically easily detected in the timestreams of TIME (see Fig.~\ref{fig:maps_collage}). We will show that this is expected given the design specifications of TIME detectors in Section~\ref{sec:reqs}. Chopping the IR-50s at 5~Hz ensures we are well into the full modulation depth regime of the sources, which is specified as 99\% modulation depth at 10~Hz. Modulation in the timestreams is therefore driven primarily by the detector and readout system response characteristics. Detectors are sampled at 100~Hz, and provide time constants that recover $>50$\% of the input amplitude at frequencies above 5~Hz ($f_{3dB} > 5$~Hz), and often have $f_{3dB}>10$~Hz. Chopping the thermal source only upon reaching each position and synchronously demodulating the output with a software lock-in amplifier allows us to avoid signal contamination in this window due to DC components (static warm mapper components in the beam) and any drifts slower than the 5~Hz chopper frequency (thermal drifts in the Hawkeye packages, end effector PCB, ambient temperature, and detector $1/f$ noise). We provide an example timeseries from a measurement campaign in Fig.~\ref{fig:maps_collage} to support the assumptions that such systematics are slowly varying or stable as the mapper holds at each position.

The PCB is screwed to a PETG raft that provides eyelets to anchor the cables. It also prevents the PCB and components from marring the mirror surfaces in the event of contact. The IR-50s are driven with an instantaneous power of $\sim$800~mW each, and are addressable via a shift register circuit in three groups: 1 in the center, 6 in an inner ring, and 6 in an outer ring. One or more groups may be powered depending on the sensitivity of the receiver and the concentration of the beam at each mirror. With all sources active, the instantaneous power draw is therefore $\sim11.7$~W, or $\sim5.8$~W on average during a 50\% duty-cycle chop sequence.

The payload mass and wiring stiffness have been minimized to relax requirements on the actuators and increase positional accuracy. The beam mapper payload has a mass of $\sim$40~g, exclusive of cabling. The cable is composed of flexible silicone-insulated 22~AWG stranded wires for the IR-50s and 28~AWG stranded ribbon cable for the serial peripheral interface (SPI) signal wires. The IR-50s are addressed via SPI to an on-board SN74HC595 shift register so that only a single pair of power wires is needed in the cable.

\subsection{Control Algorithm} \label{sec:control}

\begin{figure}
    \centering
    \includegraphics[width=0.5\textwidth]{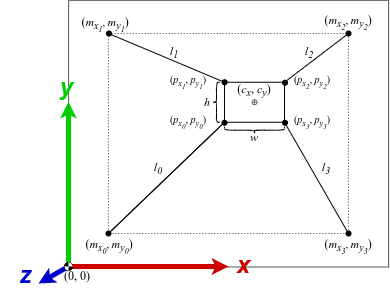}
    \caption{\label{fig:geometry} A schematic depiction of the mapper geometry. The mapper coordinate origin is in the lower left, corresponding to the intersection of the aluminum bar stock and side alignment tab. Motor mount eyelet locations are marked $m$, and are measured with calipers. Raft eyelet locations are marked $p$, and vary as a function of the mapper payload centroid, $c$, and the raft eyelet width $w$ and height $h$, which are measured with calipers. The crux of the control problem is to solve for the axis lengths $l$ of a given raft position, then translate the desired length into an angular shaft position.}
\end{figure}

The control algorithm relies on the solution to the inverse kinematic positioning problem, presented in Refs.~\citenum{williams_planar_2001, williams_planar_2003}. We illustrate this problem in Fig.~\ref{fig:geometry}. For a given payload centroid final position $(c_x, c_y)$ in the coordinate system of the mapper frame, the cable length connecting the $j$-th payload anchor point $(p_{x_j}, p_{y_j})$ to the $j$-th mapper frame anchor point $(m_{x_j}, m_{y_j})$ is
\begin{equation}
    l_j = \sqrt{(p_{x_j} - m_{x_j})^2 + (p_{y_j} - m_{y_j})^2},
\end{equation}
where the payload anchor points are positioned symmetrically about the payload centroid, at ($p_x$, $p_y$) = ($\pm\frac{w}{2}$, $\pm\frac{h}{2}$) relative to the raft centroid, where $w$ and $h$ are the raft eyelet separations. These cable lengths restrict motion to a plane in the $z$-direction and keep the rotation of the end effector about the $z$-axis aligned to the mapper axes and constant under translation. Throughout this paper, we adopt a coordinate system with the origin in the lower left corner of the mapper frame. The origin is defined as the intersection of the extruded aluminum clamping jaw (the ``lower jaw") and the side alignment tab.
The cable lengths $l_j$ are converted to final motor shaft angular positions $\theta_j$, referenced to some zero-point angular position during homing (Sec.~\ref{sec:homing}), by taking the geometry of the helical groove into account,
\begin{equation}
    \theta_j = \frac{l_j}{\sqrt{p^2 + (\pi d)^2}},
\end{equation}
where the helical groove is a thread form with pitch $p$ and minor diameter $d$. This is modeled for FDM manufacture as an ISO metric M32 $\times$ 1.5 thread. These angular positions are then converted to encoder units for commanding the motors.

In order to maintain tension on all four cables as the payload is moved to different positions in the workspace, shaft angular position changes must be synchronized. Angular velocity commands for each motor $\dot\theta_j$ are calculated from the maximum allowed motor velocity $\dot\theta_{max}$ and the angular change required by the move,
\begin{equation}
    \dot\theta_j = \frac{\theta_j - \theta_{enc,j}}{t_{move}},
\end{equation}
where $\theta_{enc,j}$ is the encoder-reported angular position at the start of the move, and $t_{move}$ is defined
\begin{equation}
    t_{move} = \frac{\max({|\theta_j - \theta_{enc,j}| : j=0,..., 3})}{\dot\theta_{max}},
\end{equation}
where $\dot\theta_{max}$ is the software-limited maximum shaft speed. This ensures that the axis with the longest distance to go always moves at the maximum allowed speed, every other axis moves at a lower velocity proportional to the distance required, and all axes reach their new required lengths at the same time. The control algorithm subdivides large moves into small segments to ensure these linear relationships hold, even for large position changes within the workspace.

By sequencing position commands, curves can be approximated and arbitrary scan patterns are possible inside the workspace. High-level control of the entire system is orchestrated by a Python program with a state machine architecture running on a laptop computer. System geometry and command profiles for each mirror are specified in human-readable comma spaced value text files. Serial commands for the motors are output via a USB-RS485 adapter. Commands for the end effector are sent as bytes over USB to a ATSAMD21 microcontroller and converted to SPI for a shift register, which switches power to the metal oxide semiconductor field-effect transistors (MOSFETs) for the IR sources. 

\subsection{Homing} \label{sec:homing}

Homing is necessary to establish the shaft position at which the length of each axis $l_j$ is zero. Since encoder feedback is available, we use it to sense this condition. Each axis is homed in sequence, with hold current to the other axes disabled. The axis being homed is driven backward, retracting cable in fixed increments until the flat side of the raft is driven against a reference surface on the integrated motor mount. Once contact is achieved, the stepper motor skips against the stop and rebounds. The encoder is queried continuously throughout this process, and the most negative encoder reading is noted as the home position. The process is repeated three times with successively smaller increments until the final estimate of the home position is known to a precision ($\sim10$~$\mu$m) that makes it subdominant to other sources of error.

\section{Performance} \label{sec:perf}

Following construction, we measure the performance of the mapper. We quantify the performance of this mapper against three metrics: (1) absolute in-plane positioning accuracy, measured relative to the mapper coordinate origin, (2) planarity, measured as the deviation from the theoretical plane being scanned, and (3) repeatability, measured by visiting the same commanded location multiple times. We derive rough requirements in Sec.~\ref{sec:reqs}, outline the test plan in Sec.~\ref{sec:testplan}, describe the measurement apparatus in Sec.~\ref{sec:opencv}, and report results in Sec.~\ref{sec:results}.

\subsection{Requirements} \label{sec:reqs}

Since our goal with this mapper is to diagnose alignment and focus issues, we would like it to be capable of movements precise enough to get information about the beam shape where it encounters each surface. With this objective in mind, we derive requirements on positioning accuracy. For simplicity, we reduce the array of IR sources to a point source (the case when using a single emitter), and see what position accuracy would be required to locate the centroid of the sensitivity pattern of a given feed at a given frequency. Since all this requires that the emitters are visible in TIME detector timestreams, we first estimate the in-band SNR.

\subsection{Visibility}\label{sec:vis}

TIME detectors are sensitive to 186-324~GHz radiation with a median channel width of $\sim2.4$~GHz. The IR-50's $\sim$920~K resistive amorphous carbon film emits unpolarized thermal photons with a peak wavelength of $\sim3$~$\mu$m, and the Rayleigh-Jeans approximation applies $\left(\frac{h \nu}{k_B T} < 2\%\right)$. The film emissivity is $\epsilon_{film} = 0.8$, so the specific intensity of the source is given in terms of the film temperature as
\begin{equation}
    I_{\nu} = \epsilon_{film} \frac{2 k_B T_{film} \nu^2}{c^2}.
\end{equation}
The radiation from the source is further attenuated due to the optical efficiency of the system $\eta_{sys}\approx0.3$\cite{hunacek_time_2018}, including reflection coefficients for mirrors, and transmission coefficients for filters, lenses, and a polarizing beamsplitter. To compute the signal level in each channel at the detector, we approximate the solid angle subtended by the source at a given mirror as the ratio of emitter area (a 1.7~mm $\times$ 1.7~mm square) to the beam area inside the FWHM, multiplied by the beam size on sky, and calculate the SNR for a fiducial circular beam FWHM on a mirror of 120~mm, which is between the smallest ($\sim$40~mm, K1) and the largest ($\sim$200~mm, F1) predicted FWHMs. Finally, we compare the flux density from the source attenuated by the efficiency terms, $S_{\nu} = 1.8$~Jy, to the TIME noise-equivalent flux density (NEFD) estimated in Ref.~\citenum{hunacek_time_2018}, NEFD($\nu$) = 0.19-0.15~Jy~s$^{1/2}$. To find the observability of an active Hawkeye in a timestream, we convert the NEFD (by definition, in one second) to the noise level in a single 100~Hz detector sample by $\textrm{NEFD}_{sample} = \textrm{NEFD} / \sqrt{0.01~\textrm{s}}$. We find that the SNR in a single sample for this fiducial beam size ranges from 0.9-1.3. We note that lock-in methods are designed to routinely measure periodic signals with SNRs $\approx1$, so the brightness of the source is well-matched to the detector sensitivity and measurement method. With the additional twelve emitters, adequate margin exists in the design to remain detectable in more diffuse beams at the cost of some spatial resolution.

\subsubsection{Absolute Accuracy} \label{sec:absacc}

In order to place a rough requirement on $xy$-plane position accuracy, we predict the receiver sensitivity patterns across each mirror using the Zemax OpticStudio model of the TIME optical system. As described in  Appendix~\ref{app:beams}, we calculate response profiles for the scanning planes above each mirror, at the shortest TIME wavelength and for a range of K-mirror positions and feedhorns. We find that the smallest beam pattern has a minor axis full width at half maximum (FWHM)$\sim$40~mm on mirror P1, so as a guideline for accuracy, the mapper would need to achieve in-plane absolute position accuracy better than $\approx$20~mm to stay within the main lobe of the smallest beam, if it had been commanded to the location of the peak.

\begin{figure}
    \centering
    \includegraphics[width=0.85\textwidth]{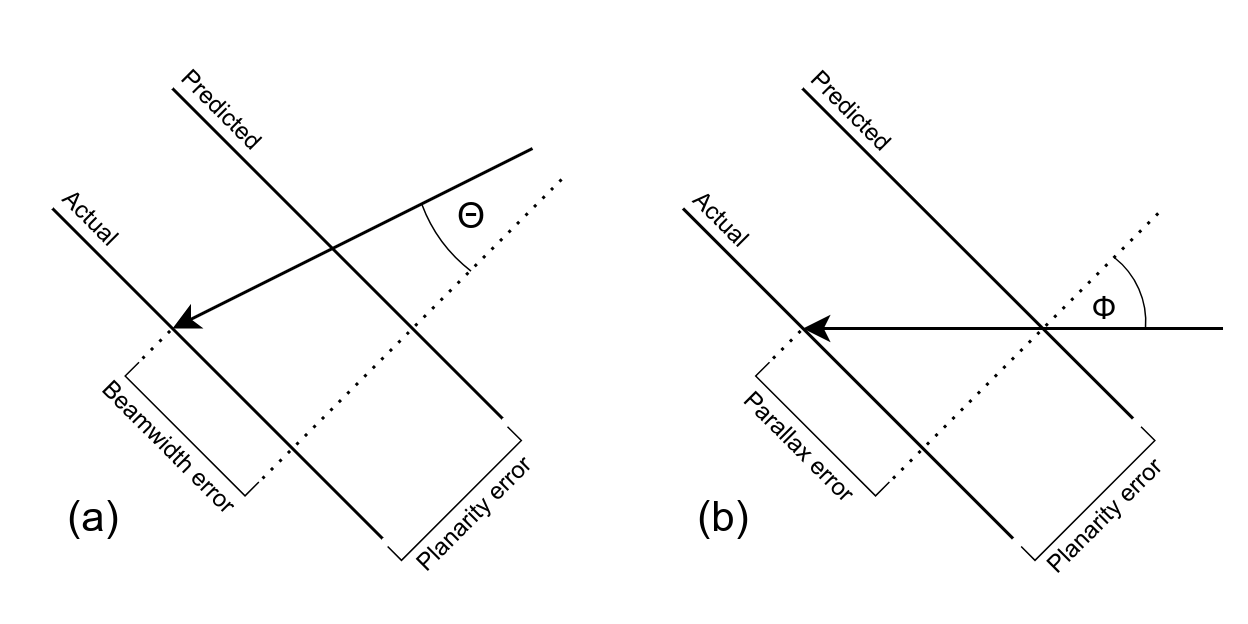}
    \caption{Schematic depiction of how systematic out-of-plane positioning errors may affect a beam mapping measurement. (a) Error in the position of the measured plane samples a converging or diverging Gaussian beam at a different beam size (beamwidth error). (b) For oblique angles of incidence, error in the position of the measured plane presents a position offset in the plane or the mirror being mapped (parallax error).}
    \label{fig:beam_errors}
\end{figure}

In order to set a requirement on the mapper $z$-direction position stability (planarity), we would like to know how motion out of the intended scan plane affects measurements of the beam pattern. Two primary effects are of concern: (1) change in main lobe size (beamwidth) as a function of Gaussian beam divergence/convergence (Fig.~\ref{fig:beam_errors}a) and (2) centroid shift due to the relative angles of the chief ray and mirror surface normal (Fig.~\ref{fig:beam_errors}b). From Zemax Gaussian beam propagation, we find that the worst-case beam divergence angle of $<$7$^{\circ}$ gives a change in main lobe radius of $\sim$0.1~mm for every 1~mm of offset in the mapper $z$-plane. From a Zemax ray trace of the chief ray, we determine that the greatest ray angle of incidence on any mirror is $\sim$55$^{\circ}$, or a parallax shift in centroid position of $\sim$1.7~mm for every 1~mm of $z$-direction offset. We neglect beam divergence going forward. 
We allocate half of the in-plane error budget to $xy$-position error and half to $z$-position error via the parallax effect. Then $xy_{req}$ = $\frac{HWHM}{2}$ = 10~mm and $z_{req}$ = 5.9~mm.

\subsubsection{Repeatability} \label{sec:repeatability}

The ability to visit the same location consistently (repeatability) gives us the flexibility to assume that revisits to a certain location (a ``tie point") will give the same detector response, allowing us to constrain drifts in the detector timeseries data as we make our maps. Imperfect repeatability then adds some ``repeatability noise" to the detector timeseries according to the gradient of the beam at the location of the tie point. We require that the inaccuracy in amplitude due to repeatability be $<$20\% of the peak amplitude. For a worst-case scenario, we evaluate at the location of highest gradient in a map. For our beam profiles, which closely follow Gaussians near the peak and into the shoulders of the distribution, the position of highest gradient is $1\sigma$ from the peak, and is $\approx 0.04$~mm$^{-1}$ for the tightest beam profile with FWHM = 40~mm. Therefore, the requirement of 20\% implies $\frac{0.20}{0.04} = xy_{req,rep}$ = 5~mm.

\subsection{Test Plan} \label{sec:testplan}

\begin{figure}
    \centering
    \includegraphics[width=0.5\textwidth, angle=270]{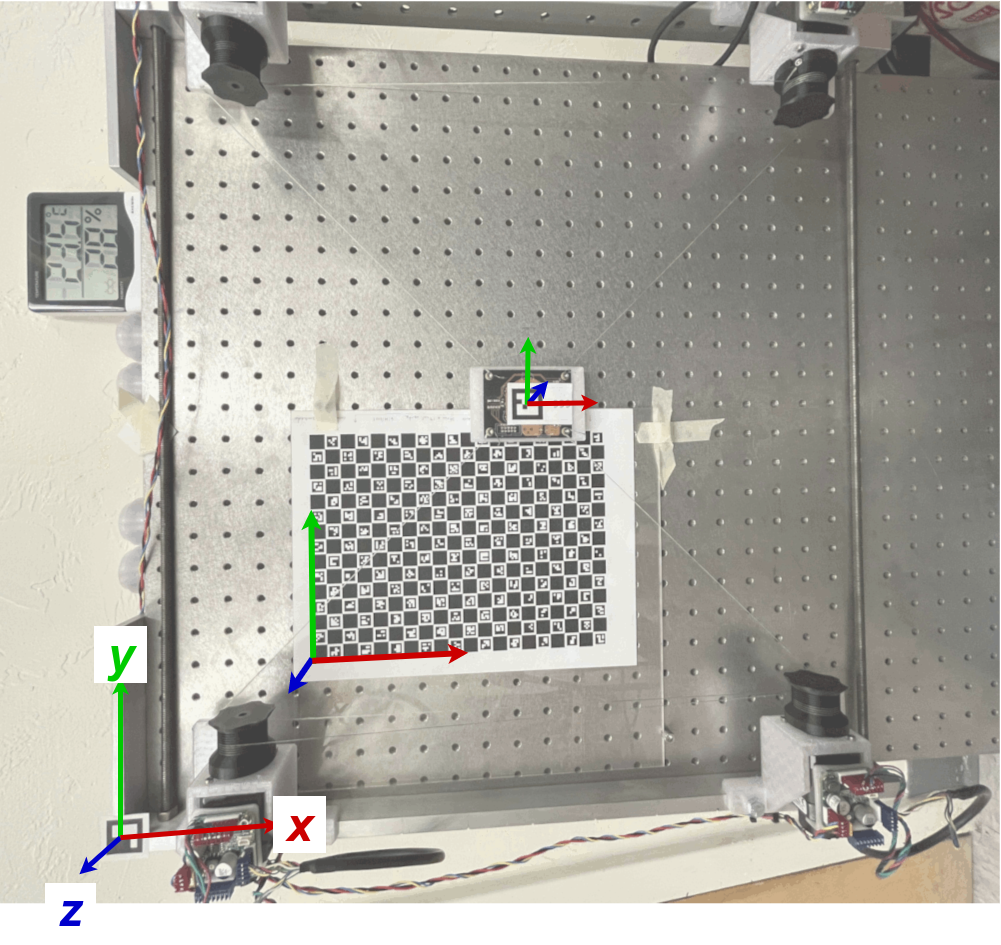}
    \caption{The mapper set up for measurement with a calibrated camera. The three features to be measured are annotated with coordinate triads. The mapper origin (lower left) is covered by a laser-printed ArUco (named for Augmented Reality, University of Cordoba) coded target. The payload is replaced by a blank printed circuit board (PCB) with another target, aligned to the center of the PCB by microscope. A laser-printed chessboard of these targets (ChArUco) bonded to a sheet of acrylic is aligned to the breadboard axes and secured flat by screws. This board establishes the axes for the coordinate system in which the other points are measured. A thermometer/hygrometer monitors the temperature and relative humidity throughout the experiment, with typical stability in the room $\pm0.1^{\circ}$~C and $\pm1\%$~RH. The gravity vector points along +$x$, down in this image.}
    \label{fig:experimental_setup}
\end{figure}

In order to measure the performance of the mapper relative to the requirements, we conduct two experiments. For each run of the mapper, it homes each axis autonomously and is fed a command profile specifying a sequence of points in a raster scan of a regular grid over the workspace. The mapper begins in the bottom left corner of Fig.~\ref{fig:error_quiver}, closest to the origin. It scans rows in +$x$ from left to right, returning to the leftmost edge to start a new row. Visits to the geometric center of the workspace are interleaved with the grid points. This command profile provides two data sets to evaluate the mapper's absolute position accuracy (grid points) and repeatability (center points). The command profile covers a 400~mm$\times$400~mm grid of 10$\times$10 points centered in the workspace, which samples performance out to the edges of the workspace while keeping reasonable density in the center. The mapper's position in 3D space is measured via a calibrated camera and custom computer vision application, explained in Sec.~\ref{sec:opencv}. During the experiments, the mapper is clamped to an optical breadboard with a known width, which allows the calculation of the eyelet positions \textit{a priori} from measurements of the motor eyelets.  The gravity vector is oriented approximately in the plane of the optical breadboard (pointing \change{down} along +$x$ in the image). The photograph in Fig.~\ref{fig:experimental_setup} illustrates the mapper in its measurement configuration.

\subsection{Position Measurement Apparatus} \label{sec:opencv}

In order to quantify the positioning accuracy of the mapper, we take photographs of the mapper with fiducial markers attached and measure their relative positions in 3D space in a post-processing step using a custom computer vision application built on the OpenCV library\cite{bradski_opencv_2000}. The fiducial markers are ArUco targets\cite{garrido-jurado_automatic_2014} laser printed on an adhesive paper backing, which we attach to the end effector and the mapper origin. A laser-printed grid of ArUco targets (named for Augmented Reality, University of Cordoba) interleaved with a chessboard pattern (a so-called ``ChArUco'' board) is bonded to an acrylic panel and fixed flat against the optical breadboard, providing a reference set of axes aligned with the mapper coordinate frame as described in Sec.~\ref{sec:frame}. A ChArUco board is used instead of a chessboard pattern because the interspersed coded targets can be identified in images even when some are partially obscured by the mapper, allowing reconstruction of all chessboard coordinates. With these three fiducials, we can measure the position of the mapper in its native coordinate frame.

As the mapper executes the command profile, photographs are taken at each position. At each position, we observed periodic out-of-plane motion (along the $z$-axis) of the payload that required time to settle, for which we allowed $\sim$1-2 seconds. Photographs are taken from a tripod-mounted camera with a remotely-triggered shutter. An iPhone 13 mini, with a locked focus and without image stabilization, is sufficient for our purposes once the lens distortion parameters are calibrated. Calibration is performed using Zhang's method\cite{zhang_flexible_2000}. Target positions in the mapper coordinate frame are estimated from the intrinsic size and image position\cite{marchand_pose_2016} using OpenCV. See Table \ref{tab:camera} and Appendix~\ref{app:cal} for additional details.
\begin{table}
\begin{tabularx}{\textwidth}{l|l|l|l|X}
    \textbf{Category} & \textbf{Parameter} & \textbf{Value} & \textbf{Units} & \textbf{Comments}\\
    \hline
    \multirow{10}{4em}{Optical Parameters}
    & Focal length (35mm equiv.) & 14 & mm & \raggedright\arraybackslash Approx. (manufacturer spec.) \\
    & Aperture & f/2.4 & -- & Fixed \\
    & Vertical resolution & 4032 & px & -- \\
    & Horizontal resolution & 3024 & px & -- \\
    & Vertical field of view$^*$ & 101.218 & deg & -- \\
    & Horizontal field of view$^*$ & 84.916 & deg & -- \\
    & Vertical focal length$^*$ & 1.655 & mm & -- \\
    & Horizontal focal length$^*$ & 1.652 & mm & -- \\
    & Pixel size & 1.0 & $\mu\textrm{m}$ & \raggedright\arraybackslash Approx. (manufacturer spec.) \\
    & Calibrated reprojection error & 0.638 & px & RMSE \\
    \hline
    \multirow{4}{4em}{Capture Settings}
    & Shutter Speed & 1/61 & sec & Fixed \\ 
    & ISO & 320 & -- & Fixed \\
    & Image format & JPEG & -- & -- \\
    & Automatic lens correction & Disabled & -- & iOS setting \\
    \hline
    \multirow{2}{4em}{Geometry}
    & Working distance & 0.3 & m & \raggedright\arraybackslash Approx. (varies across scene) \\
    & ArUco target size & 24.93$\pm$0.01 & mm & -- \\ 
    \hline
    Distortion Model & Pinhole & -- & -- & 6 radial, 2 tangential distortion \\
    \hline
\end{tabularx}
\caption{Parameters of the camera setup used for mapper position measurement. Parameters marked with $*$ are derived from the camera calibration matrix. Complete pinhole camera model parameters and distortion parameters are given in Appendix~\ref{app:cal}.}
\label{tab:camera}
\end{table}

\subsection{Results} \label{sec:results}

Here we present results from measuring 200 images: the mapper performs a 100-point raster scan in the $xy$ plane, interleaved with 100 visits to the center of the workspace, as discussed in Sec.~\ref{sec:testplan}.

\subsubsection{Absolute Accuracy} \label{sec:acc}

\begin{figure}
    \centering
    \includegraphics[width=0.75\textwidth]{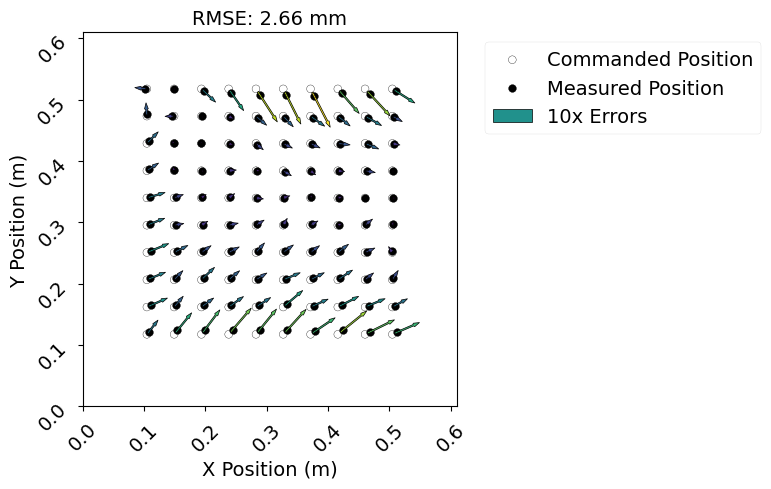}
    \caption{A 2D quiver plot of the mapper's absolute position errors, projected into the $xy$-plane. Open circles denote commanded positions, filled circles denote measured positions, and filled arrows illustrate the direction and 10 times the magnitude of the errors. The RMSE in the $xy$-plane is calculated to be 2.7~mm. The first position is in the lower left, and the scan proceeds from left to right in rows of increasing $y$.}
    \label{fig:error_quiver}
\end{figure}

The absolute accuracy of the mapper, defined as the position relative to the mapper origin, is shown in Fig.~\ref{fig:error_quiver}. Note that the error magnitude is multiplied by 10 for display. The dominant error is a sag in the $+x$ direction, following the gravity vector, which also has a small $-z$ component due to the tilt of the breadboard ($\sim7$~deg). Taking the root-mean-square-error (RMSE) of all points in the $xy$ plane gives us an overall position accuracy of RMSE = 2.7~mm.

\begin{figure}
    \centering
    \includegraphics[width=\textwidth]{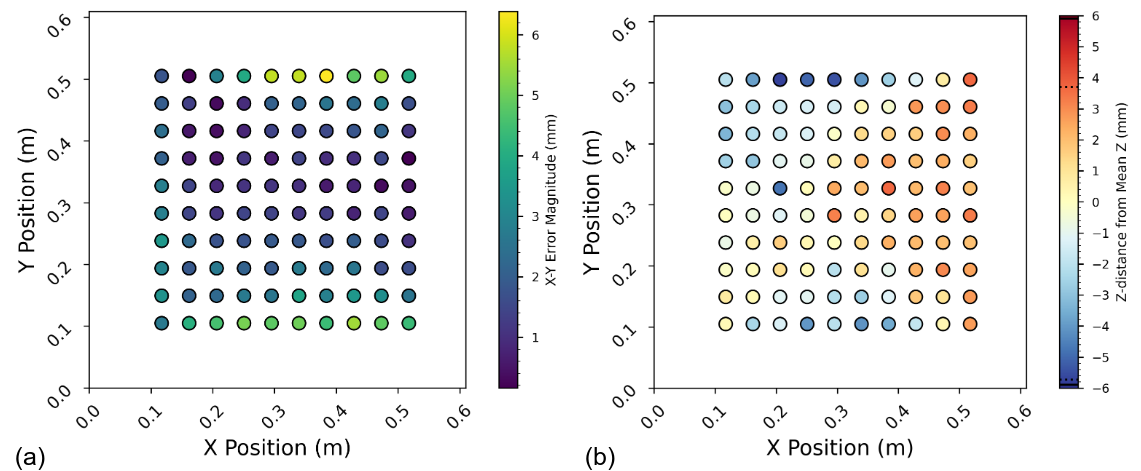}
    \caption{(a) 2D map of $xy$ plane errors, from the same scan as Fig.~\ref{fig:error_quiver}. Error magnitudes are the length of the error vectors in the $xy$ plane. The max error magnitude is 6.4~mm. (b) 2D map of $z$-deviations from the effective scan plane, defined as the mean $z$-coordinate. Planarity requirements $z_{req}$ and $z_{goal}$ are judged relative to this effective scan plane. The largest positive and negative excursions (indicated by dotted lines in the colorbar) are 3.7~mm and -5.7~mm, respectively.}
    \label{fig:coarse_error_pos}
\end{figure}

In Fig.~\ref{fig:coarse_error_pos}, we show two 2D error maps: one of $xy$-plane error magnitudes, and one of $z$-axis deviation values. In Fig.~\ref{fig:coarse_error_pos}a, we find a maximum error of 6.4~mm. In Fig.~\ref{fig:coarse_error_pos}b, we plot the deviation in $z$-position relative to the mean. We find a greatest deviation of -5.7~mm. It is the deviation from this effective scan plane that we judge against the planarity requirement.

\begin{figure}
    \centering
    \includegraphics[width=.75\textwidth]{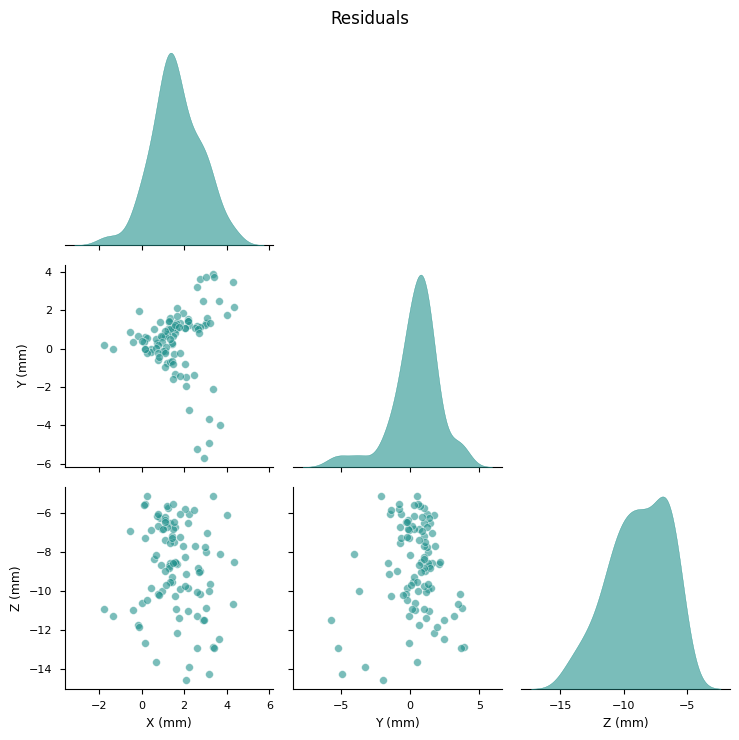}
    \caption{Corner plot of $x$-, $y$-, and $z$-position residuals, the measured minus commanded positions, expressed in the coordinate frame of the mapper origin.}
    \label{fig:coarse_error_corner}
\end{figure}

Figure~\ref{fig:coarse_error_corner} displays the $x$-, $y$-, and $z$-position residuals in the coordinate frame of the mapper simultaneously, to allow investigation of any correlations between them. The residuals are non-Gaussian with fat tails, but with no obvious correlations.

\subsubsection{Repeatability} \label{sec:repeat}

\begin{figure}
    \centering
    \includegraphics[width=0.75\textwidth]{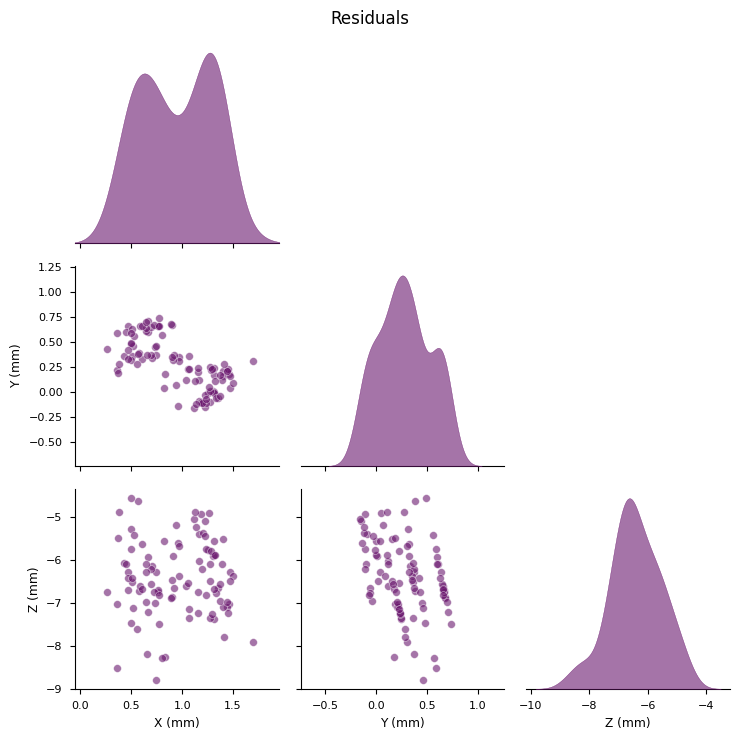}
    \caption{Corner plot of repeatability experiment. Each point is one measurement of the position of the mapper after being commanded to drive to the center of the workspace. From these plots, we observe bimodality in the $xy$-plane, and striping and correlations in the $yz$-plane, so we choose to describe the repeatability errors using the 16th-84th percentile range of the residuals (quoted in the text), rather than parameters of a Gaussian distribution.}
    \label{fig:repeatability_corner}
\end{figure}

The repeatability of the mapper is measured from images of each visit to the center of the workspace using the same technique described in Sec.~\ref{sec:opencv}. Visits of this position are interleaved with the grid points in the raster scan, which is intended to avoid correlation of the position error with the previous position, as might occur if we had moved back and forth repeatedly between two positions. Figure~\ref{fig:repeatability_corner} shows the $x$, $y$, and $z$-plane projections of the commanded and measured positions in the coordinate frame of the mapper. All points in the repeatability dataset are contained by $\Delta x\leq$ 1.4~mm, $\Delta y\leq$ 0.9~mm, and $\Delta z\leq$ 4.2~mm. Since the distributions are highly non-Gaussian, we report half the spread of the 16th and 84th percentile values for rough comparisons: position repeatability is achieved with $\delta x$ = 0.41~mm, $\delta y$ = 0.30~mm, and $\delta z$ = 0.85~mm.

\section{Discussion} \label{sec:disc}

\subsection{Sources of Error} \label{sec:errsrcs}

\subsubsection{Systematics and Measurement System Errors}

In evaluating performance, we consider both errors in the mapper system and the measurement system. Human error in target placement can lead to a systematic bias in measured position. The targets were positioned on the payload and mapper to better than 0.1~mm in $x$ and $y$, which we will show are subdominant to the camera measurement error.

For each target in the scene (origin and payload), the image-space corner locations are identified using the OpenCV implementation of the method from Ref.~\citenum{garrido-jurado_automatic_2014} and refined to subpixel accuracy using the method from Ref.~\citenum{wang_apriltag_2016} before being passed to the Perspective-N-Point algorithm that solves for the position relative to the camera\cite{fischler_random_1981}. The accuracy of corner refinement depends on the gradient of the marker pattern in the image (i.e., contrast) and image sensor noise. Undertaking a comprehensive characterization of corner refinement performance is beyond the scope of this work, however we note that spot-checks of the image histograms of target regions yield estimated contrasts of $\sim$4:1 and SNR of $\sim$5, so we expect satisfactory performance. We assume subpixel refinement succeeds in estimating positions to a fraction of a pixel size ($\ll$ 1~px, or $\ll$ 0.3~mm at our working distance) and neglect its impact.

\begin{figure}
    \centering
    \includegraphics[width=0.75\textwidth]{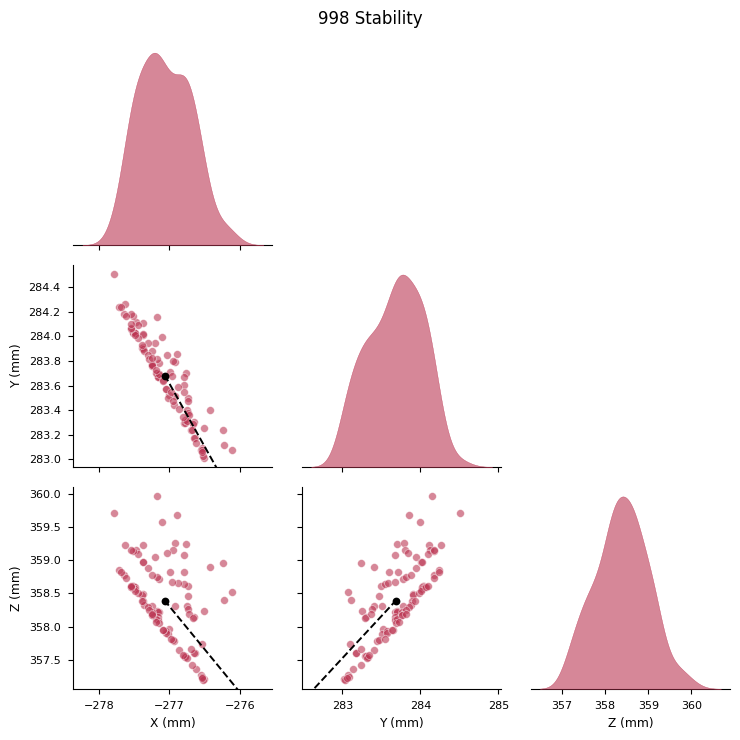}
    \caption{Corner plot of $x$-, $y$-, and $z$-positions of the ArUco target affixed to the mapper frame origin, this time in the coordinate frame of the camera: $z$ outward perpendicular to the sensor, $x$ along sensor long dimension, and $y$ completing the right-handed coordinate system. Taking the raw target positions in the camera coordinate frame allows us to see the camera reconstruction errors for repeat measurements of a single target. On the scatter plots, a dashed line extends from the mean position to the camera origin, indicating the line of sight from camera to target.}
    \label{fig:origin_dist}
\end{figure}

From camera calibration, we obtained an RMS reprojection error of 0.638~px; this quantifies the errors that remain after deriving the best-fit intrinsic parameters of our camera (focal lengths, distortion parameters, principal image point, etc.). If camera calibration is successful, reprojection errors are approximately circular Gaussian in image space, but the 3D position uncertainty ellipsoid of points de-projected back into 3D space may be far from spherical, and in general varies with target pose. We assume the mapper origin remains fixed in world-space, and show the measured position distribution in Fig.~\ref{fig:origin_dist}. One caveat of this measurement is that this probes one point in the error volume: the error distribution will change shape as a function of position in the measurement volume, a consequence of the covariance matrix of the error ellipsoid being a complicated function of the camera parameters themselves, the camera model assumed, the field position, orientation relative to the target, and working distance. See Ref.~\citenum{pentenrieder_analysis_2007} (especially Fig. 5) for a full simulation-based treatment of this type of error, which conveniently covers a similar target size and working distance to our case.

To inform a conservative estimate of the errors of the camera measurement system, we calculate the approximate line of sight distance to this target as $r_o \approx 500$~mm. The dominant error component is along the camera line of sight, as shown by the alignment of the residuals along the line of sight (dashed line) in Fig.~\ref{fig:origin_dist}. Due to the measurement geometry, errors in image-space target locations are translated into distance errors via scaling by the true target size. The magnitude is $\delta_{LOS}$ = 0.7~mm. With error magnitude and working distance, we estimate the accuracy of the camera system in terms of millimeters of error per millimeter of distance by assuming a spherical Gaussian error distribution with this value and adopting a conservative floor on the error at $\delta_{min}$ = 0.100~mm:
\begin{equation}
    \delta_{abs} = \delta_{min} + \frac{\delta_{LOS}}{r_o} r = 0.100~\textrm{mm} + 0.001 \, r \, ~\textrm{mm mm}^{-1},
\end{equation}
at a distance $r$ from the camera. We expect similar performance across the field of view, with degraded accuracy with increasing distance, consistent with previous studies such as Ref.~\citenum{pentenrieder_analysis_2007}. We note that this is an overestimate of the $x$-$y$ errors imparted by the measurement system in this case, since the camera line of sight is nearly perpendicular to the $xy$ plane across much of the mapper workspace.

\subsubsection{Spooling Error} \label{sec:error_spooling}

Several assumptions, mentioned in Sec.~\ref{sec:desc}, were made in designing the mapper and writing the control software, including: no cable stretch, negligible homing error, rigid motor mounts, negligible motor angular position error, and linear cable length spooled for linear motor angle changes. Of these, the greatest impact on performance is likely to come from the last.

\begin{figure}
    \centering
    \includegraphics[width=0.5\textwidth]{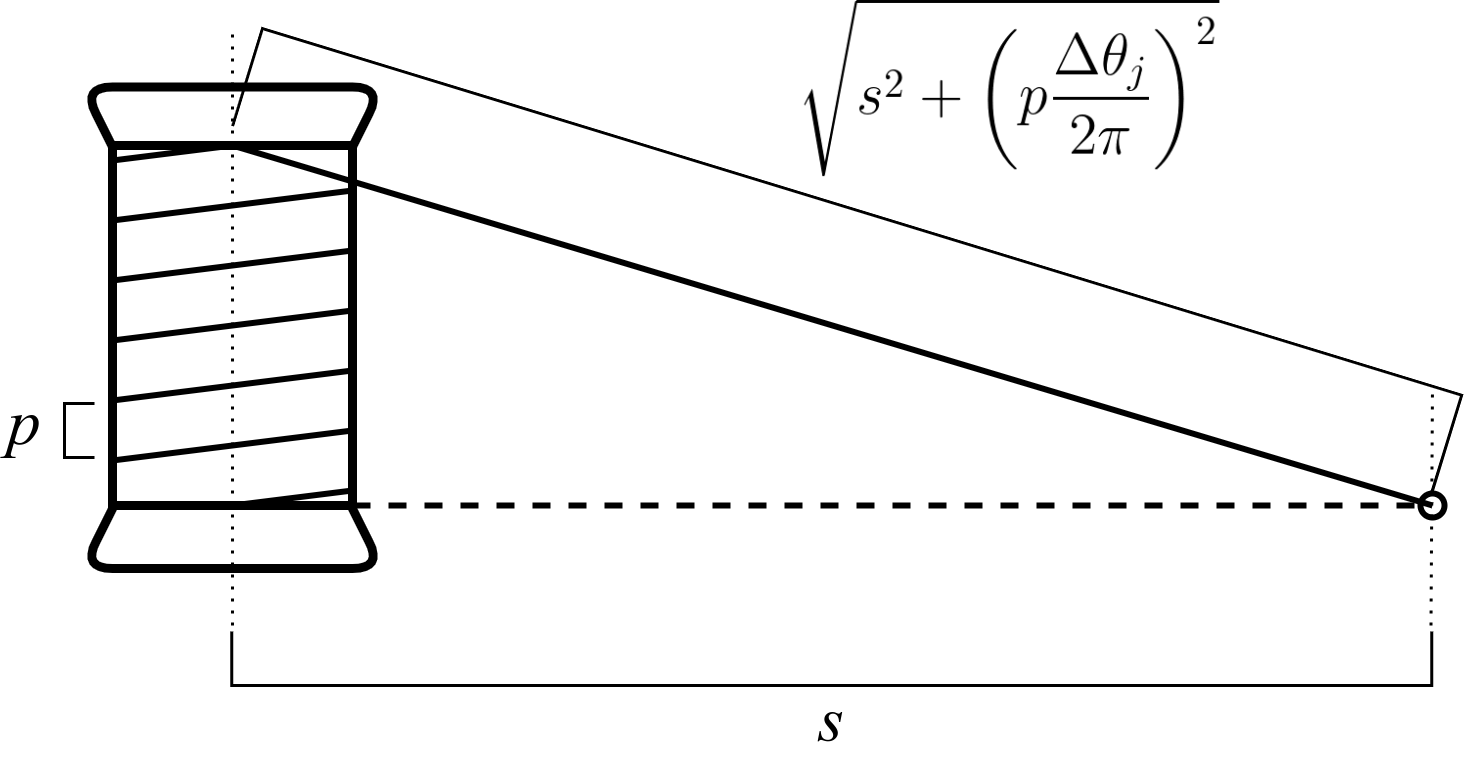}
    \caption{A schematic depiction of geometry unaccounted for in the control algorithm described in Sec.~\ref{sec:control}. For drum pitches $p$ or angular changes $\Delta \theta_j$ sufficiently small, or separations of drum and eyelet $s$ sufficiently large, the extra length of the hypotenuse between eyelet and drum becomes negligible.}
    \label{fig:spoolingerror}
\end{figure}

By routing the cable from each drum through a distant eyelet on another motor mount, we reduce but do not fully eliminate some nonlinearity in the length of cable unspooled for a given change in drum angle. This nonlinearity arises because the cable exit position along the drum changes with angle. If the cable exits the helical groove perpendicular to the axis of rotation when no cable is wound on, after many turns of the drum, the angle and height of the cable as it exits the groove has increased relative to the initial position, increasing the effective span between the drum and the eyelet. We illustrate this in Fig.~\ref{fig:spoolingerror}.

The error in cable length change incurred by approximating this span as constant is $\sim 0.14$~mm, for approximate values for the mapper $s \approx$ ~500 mm, $\Delta\theta_j =$ 16$\pi$, and $p =$ 1.5~mm. This is the maximum excess cable that each axis could possibly contribute, and this position-dependent excess length could lead to noticeable loss of tension at some positions in the workspace. Detailed modeling of the impact on mapper position due to these cable length errors and gravity is outside the current scope, but the model of Fig.~\ref{fig:spoolingerror} provides a plausible explanation for two effects we see in the data. First, in Fig.~\ref{fig:error_quiver}, we see that areas of the workspace near the edges appear to be more affected by gravity (pulled toward +$x$) and have higher position errors than those near the center. From Fig.~\ref{fig:spoolingerror}, we know that the accrued length error in an axis is proportional to the payload's distance from the eyelet, therefore the workspace location that minimizes the joint error from all axes is the center. We also know that the length errors we have described will always lead to some loss of tension, because cable length errors are always positive for this homing scheme.

\subsection{Requirements Verification} \label{sec:reqsver}

From our treatment of sources of error, we have confidence that our measurement setup is capable of determining whether the mapper meets the requirements set forth in Sec.~\ref{sec:reqs}. The primary source of measurement error, the camera measurement, contributes $\leq$0.7~mm in each axis. Since we use the difference of 3D positions to find each mapper position relative to the origin, the final measurement uncertainty by standard error propagation is $\sigma$ = 1.0~mm in each axis. We conclude that our measurement errors are subdominant to the positioning errors, with in-plane RMSE = 2.7~mm, in-plane max error 6.4~mm, and out-of-plane max error -5.7~mm. The measurement errors are similar in magnitude to the repeatability errors (0.41~mm, 0.30~mm, and 0.85~mm). We have also identified and characterized, through the error maps, a source of algorithmic error arising from the geometry of the mapper, which accounts for the position errors we see at the edges of the workspace. Most importantly, we have shown that this kind of error can be compensated with a control algorithm update that compensates for the error of Fig.~\ref{fig:spoolingerror} into the calculation of spool angles for a given position.

Now, we judge performance against the requirements set forth in Sec.~\ref{sec:reqs}. A requirement is only judged a pass if the measured error is less than the requirement over the entire workspace that was measured. With this established, even without the algorithm corrections, we judge that the mapper meets the accuracy and repeatability requirements across the entire workspace (see Sec.~\ref{sec:results}, Fig.~\ref{fig:coarse_error_pos}), with the caveat that these measurements were taken in only one orientation in line with the gravity vector, which most closely approximates the scenarios of mirrors K2 and P1. We summarize the performance in Table~\ref{tab:reqs}. 

\begin{table}
\begin{tabularx}{\textwidth}{l|l|l|l}
    \textbf{Item} & \textbf{Req. (mm)} & \textbf{Value (mm)} & \textbf{Result} \\
    \hline
    $xy_{req,rep}$ & 5.0 & 0.41 & PASS \\
    $xy_{req}$ & 10 & 6.4 & PASS \\
    $z_{req}$ & 5.9 & 5.7 & PASS \\
    \hline
\end{tabularx}
\caption{Summary of mapper positioning accuracy against the accuracy requirements. For $xy_{req,rep}$, we quote the worst of the $x$ and $y$ components of the repeatability error, while for $xy_{req}$ and $z_{req}$, we quote the worst errors measured inside the workspace. We assess that the mapper meets the requirements.}
\label{tab:reqs}
\end{table}

\subsection{Real World Performance}

During the 2024-2025 and 2025-2026 seasons, the TIME instrument was again deployed on the ARO 12~m telescope. Each year, the initial alignment of the warm optics and cryostat were set with a laser tracker. The components were installed within 1~mm of the design positions and 0.1~deg of the design angles, meeting tolerances. We present a selection of 2026 beam maps to illustrate the mapper data product. In 2026, each mirror in the optical system was mapped over the course of two days: P1 and F1 were mapped on January 23, 2026, and K1, K2, and K3 were mapped on January 24. We achieved mapping times of 42 minutes per mirror over a 25 $\times$ 25 grid in a snaking raster pattern, for an average speed of $\sim$15 points per minute. For the K1 maps presented here, only a single emitter was used. K2 and K3 used 7 emitters, while F1 and P1 employed all 13.

\begin{figure}
    \centering
    \includegraphics[width=0.4\textwidth]{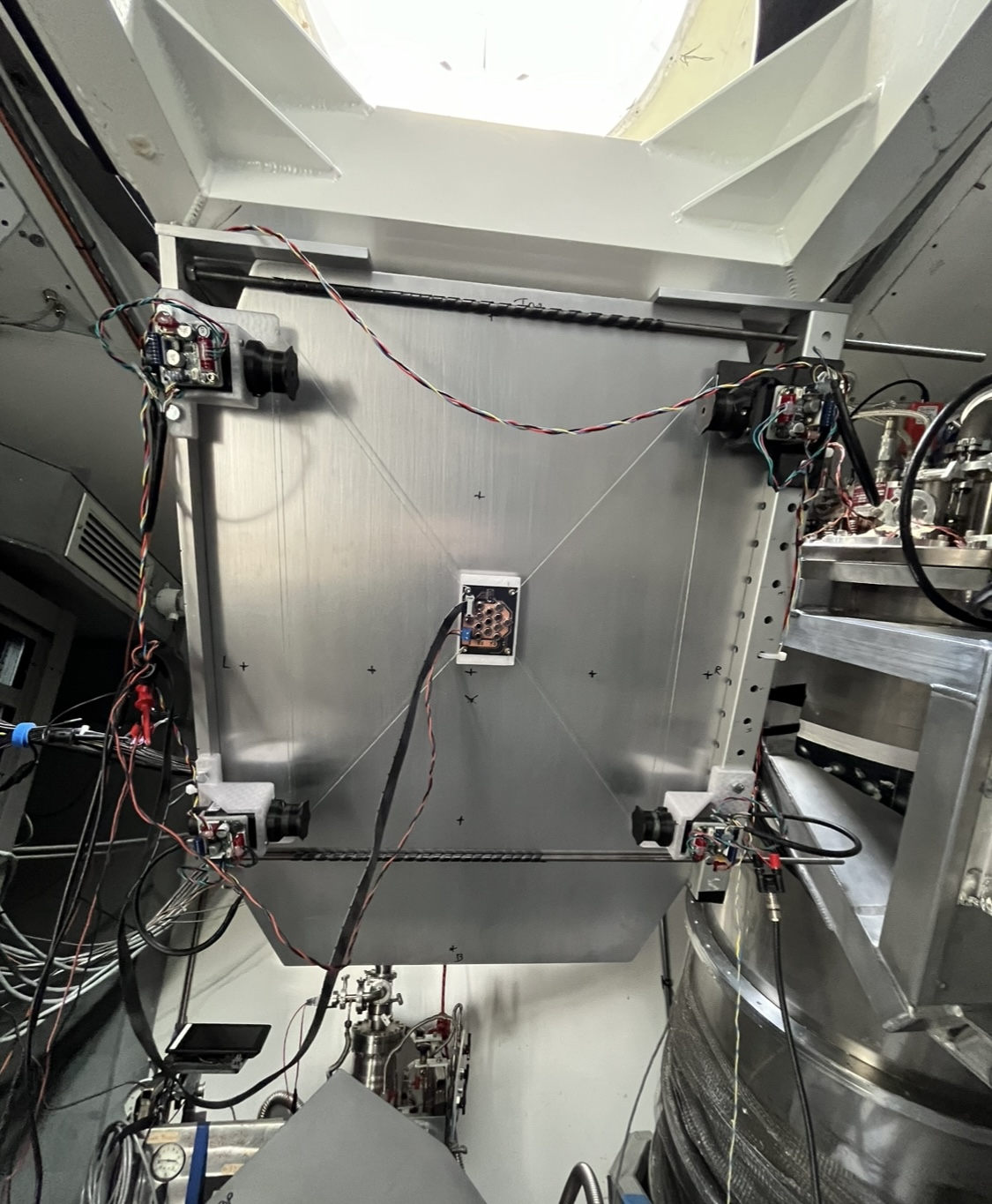}
    \caption{The mirror mapper mounted on K3 in 2025.}
    \label{fig:setup_k3}
\end{figure}

\begin{figure}
    \centering
    \includegraphics[width=\textwidth]{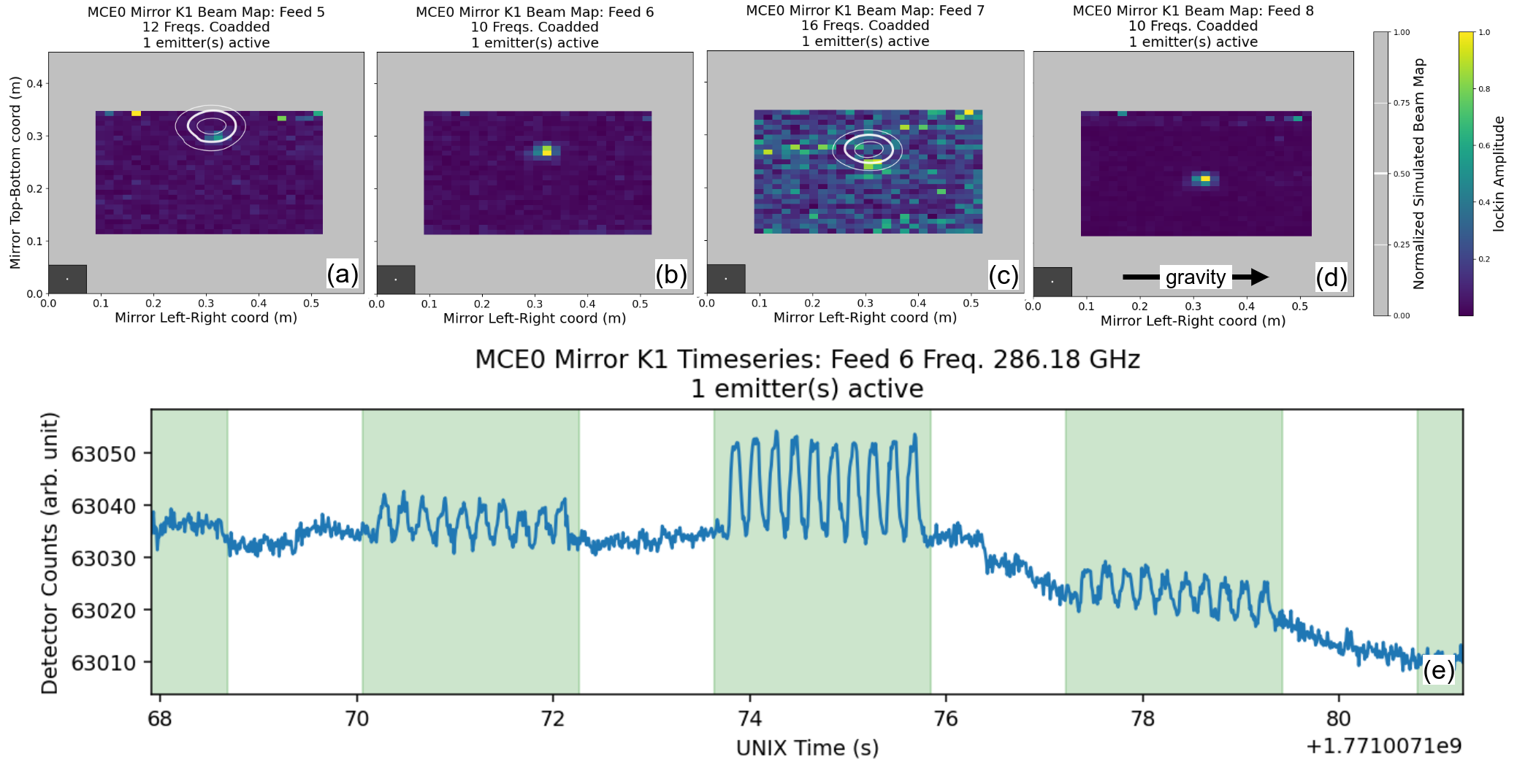}
    \caption{(a-d) Selected maps for four spatial feeds in one polarization. \change{Pixels are 16.4~mm by 8.8~mm.} The silver region represents the actual size of the mirror - the mapper does not have full coverage of each mirror. The size of the PCB is denoted by the small grey box in the corner. The single \change{1.7~mm $\times$ 1.7~mm} emitter, denoted with a white square, is  small compared to the map pixel size, so the Zemax beam profile is not convolved with the emitter spatial pattern. Maps show the position-dependent sensitivity of the detectors to the chopped 5~Hz emitter signal. The predicted beam pattern contours of feeds 5 and 7 are overplotted in white; the Zemax simulation omits every other feed for processing speed. The beam centroids move across the mirror in the spatial sequence as expected. The mapped pattern is smaller than predicted, indicating that a cryostat axial position offset places the focus near K1 closer to the mirror than intended, which can be compensated with the telescope secondary mirror actuator. (d) The direction of the gravity vector projected into the mapper plane is indicated. (e) A segment of a single detector timestream that contributed to the map in (b). The green windows highlight the $\sim2$~s \change{intervals} when the raft is stationary at each location while the emitter \change{is chopped on and off}. The window in the center of the plot corresponds to the brightest pixel in (b). We note that the baseline is stable across all windows shown, indicating thermal and detector drifts on this timescale are small relative to the emitter signal.}
    \label{fig:maps_collage}
\end{figure}

In Fig.~\ref{fig:maps_collage}, we show a sequence of maps of K1, illustrating a distinct beam pattern moving across the mirror as a function of spatial feed position. Detector timeseries from multiple receiver frequencies are co-added to increase signal to noise ratio. The 5~Hz chopped source amplitude at each position is extracted from these timeseries using the complex lock-in method, implemented in digital signal post-processing. We mask out detector samples with large (6 $\times$ RMSE) time-domain impulse spikes that could bleed power into the 5~Hz signal. Each map is then normalized to a peak amplitude of 1, and contours of the normalized predicted beam pattern for the corresponding spatial feed are overplotted in white for comparison. This method of coarse comparison allows us to evaluate the beam centroid positions and FWHMs for alignment and large detect large aberrations, \change{which is the primary purpose of the system. More detailed mapping could also be performed by, for example, scanning on grids that push closer to the RMSE position error and/or using smaller emitter elements to better sample beams near foci, or by making long-integration maps with a larger number of emitters active to recover low-level beam structure.}

The beam pattern for K1 is expected to be compact, as it is near the Cassegrain focus, but the patterns we see in Fig.~\ref{fig:maps_collage} are smaller than expected. We interpret this as an axial position offset of the feedhorn locations inside the cryostat that places the focus closer to the surface of K1 than intended. Routine compensation with the telescope secondary actuator still allows TIME to reach focus on-sky. Centroid positions are roughly as expected, with small position offsets. We have shown that the mapper is capable of rapid measurement of beam profiles on each optic, facilitating useful comparison with simulated data and helping to narrow down the problem space.

\section{Conclusions} \label{sec:conclusions}

In this work, we presented a novel design for a beam mapper consisting of a bank of thermal sources positioned by a reconfigurable planar cable-driven robot. We gave a brief overview of the field of parallel robotics and the TIME concept of operations. We show that a simple, low-cost design using off the shelf aluminum extrusions, FDM 3D-printed components, and commercially available stepper motors and drivers is feasible for low-mass, low tension payloads. The design naturally permits scaling to larger workspaces, needing only a larger mapper frame, longer cables, and configuration file updates to be useful. We describe the operation of an apparatus for measuring the positioning accuracy of the robot in 3D, using a widely available consumer camera, and present the analysis methods and results of position measurement. We determine that the mapper has better than required position accuracy for our application. \change{The achievable map resolution is limited by the positioning accuracy, for which we measure an RMSE$\sim$2.7~mm across the workspace, allowing us to sample all beam planes in our system with at least two pixels per beam FWHM.}

We performed the algorithm update described in this work and fielded the system in the 2024-2025 and 2025-2026 seasons.  Further work could refine the homing process to include physical or capacitive sensor inputs (limit switches) for the EZStepper driver boards to directly sense when homing is achieved, use more advanced driver boards with current-sensing capability for sensorless homing, or employ methods of auto-calibrating the cable lengths that have shown promise \cite{borgstrom_nims-pl_2009, miermeister_auto-calibration_2012, gosselin_initial_2018}, but add complexity. Although reflections that can be traced to the mapper have not been observed, standard practice to avoid all reflections would involve covering exposed surfaces in an absorber material, such as carbon-loaded foam. Such surfaces would appear as $\sim$300~K blackbodies rather than the less emissive aluminum of the mirrors. Brighter or larger-area sources could increase signal strength, but to our knowledge, the variety of fast, pulsable IR emitter sources is small, and these exact sources have been used in mm- and far-IR domains successfully\cite{hunacek_time_2018, wilson_toltec_2020}. Future optical component designs could incorporate additional mechanical references in the form of pin holes and fastener clearances for repeatable and secure mating, and design in keep-out zones around the edges of optics to better accommodate in-situ beam mapping equipment. Stereolithography files are available for the FDM components, and the optical simulation, control software, and OpenCV measurement code are available with permissive open source licenses to promote reuse. Looking beyond the TIME experiment, the mapper we describe here can easily be reused in any lab environment for mapping beams exiting the cryostat windows of other experiments by clamping it to a simple open frame of a known size.

\appendix

\section{Beam Map Predictions} \label{app:beams}

In this section, we present beam map simulations that inform the mapper requirements. We present beam maps for the smallest wavelength $\lambda$ = 922~$\mu$m, and mirror closest to the Cassegrain focus, K1, as an example of what is done to compare the optical model to measured beam maps. Simulations for each mirror are performed in the time-reversed sense (a radiating beam launched from the cryostat outward toward the sky), and proceed in a leapfrog fashion, originating from the Zemax-calculated POP illumination and phase pattern of the preceding mirror.

\subsection{Numerical Calculations}

In order to predict the sensitivity patterns at each surface, we use the Zemax model of the TIME optical system. We start with Zemax OpticStudio physical optics propagation (POP) reports at each surface for a beam launched in the time-reversed sense from each feedhorn. Zemax provides the complex electromagnetic field data (amplitude and phase maps) at each surface. These maps are defined in a plane centered on and perpendicular to the chief ray, with some rotation about it, and are provided after Zemax has applied a phase transfer matrix, which propagates the beam from the input to the output side of the optic by accounting for the curvature of the optic and its orientation in the system. These data are however insufficient to understand the illumination of each point on each mirror surface itself. To calculate the illumination patterns, we rotate the POP amplitude and phase data from the Zemax output files into the global Zemax coordinate system, and propagate the amplitude and phase leaving the previous surface to a grid of points on the surface of interest. The relevant operation for propagation in the near-field region is the brute-force solution of the Rayleigh-Sommerfeld integral,
\begin{equation}
    E(x, y, z) = \frac{1}{i \lambda} \iint_{-\infty}^{\infty} E(x', y', z') \frac{e^{ikr}}{r} dx' dy',
\end{equation}
where $i$ is the imaginary unit, $\lambda$ is the wavelength, $E$ is the value of the complex electromagnetic field at source coordinates ($x'$, $y'$, $z'$) and field coordinates ($x$, $y$, $z$), $k = 2 \pi / \lambda$ is the wavenumber, and
\begin{equation}
    r = \sqrt{(x' - x)^2 + (y' - y)^2 + (z' - z)^2}
\end{equation}
is the distance between a pair of source and field points. We calculate the integral for each field point by brute force, rather than using a Fourier transform-based method, in order to allow calculating the integral across a destination surface that is not parallel to the source plane or plane of propagation. We approximate the value of this integral at each field point as a discrete sum of contributions from each source point in the illumination pattern of the previous optic. Our source and detectors are incoherent, so we present magnitude-only plots, although phase information is available from the calculation.

\begin{figure}
    \centering
    \includegraphics[width=\textwidth]{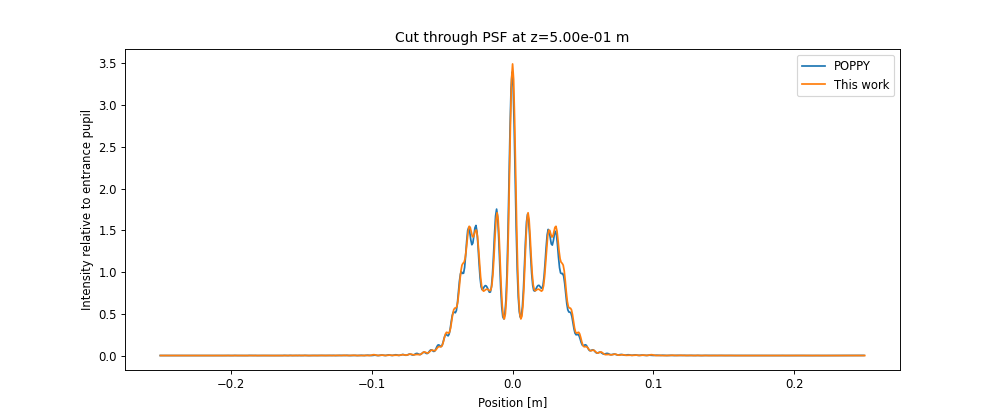}
    \caption{Comparison between our method and POPPY, showing the near-field Fresnel interference pattern of 922~$\mu$m light after impinging on a $\sim$10~cm aperture and propagating 50~cm.}
    \label{fig:poppy_comparison}
\end{figure}

We validate this approach by comparison to the angular spectrum method from the \texttt{POPPY}\cite{perrin_poppy_2016} physical optics package for a case where they should agree, plane-parallel geometry. See Fig.~\ref{fig:poppy_comparison}.

\subsection{Predicted Amplitude}

\begin{figure}
    \centering
    \includegraphics[width=.8\textwidth]{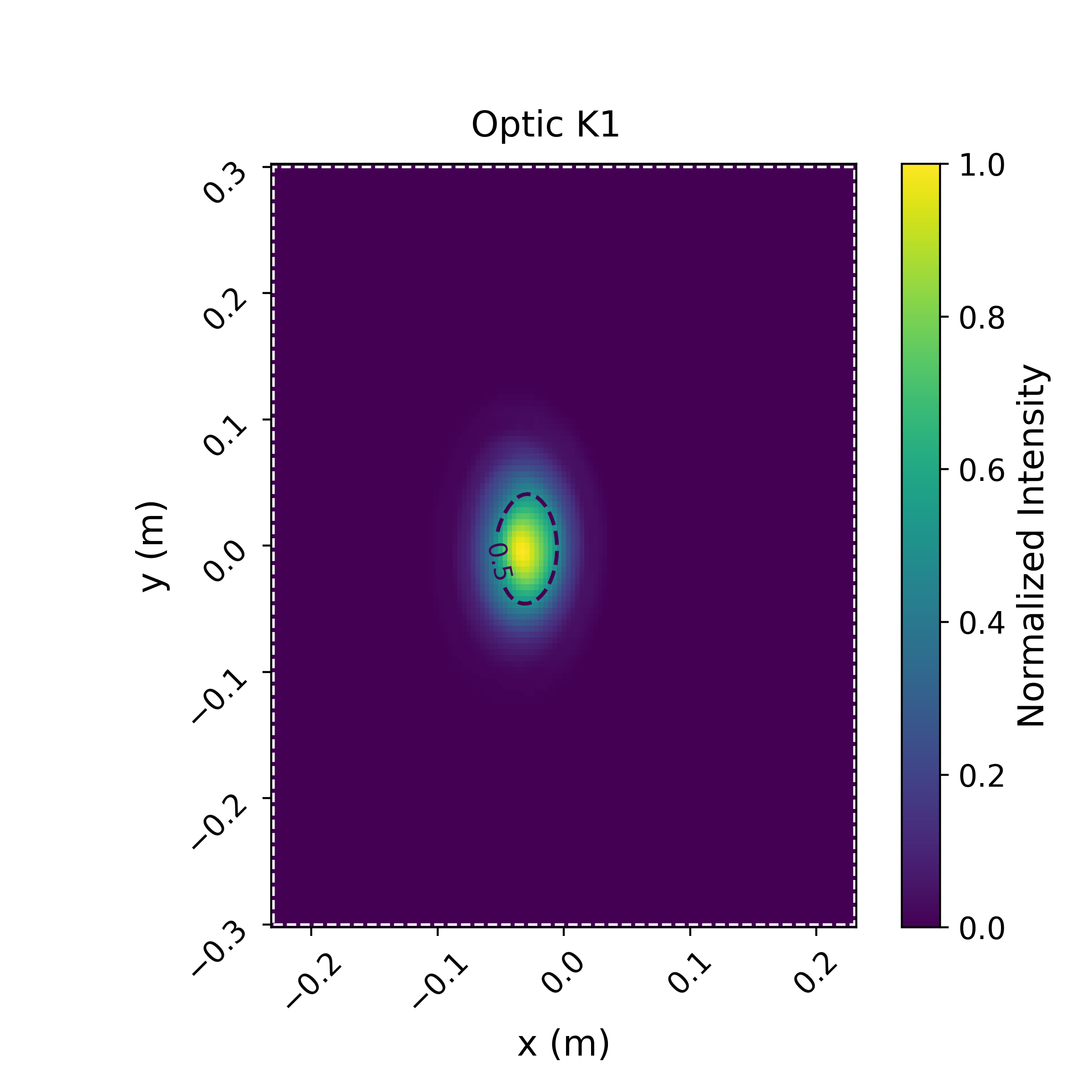}
    \caption{Simulated illumination of K1, K-mirror in neutral position, feed 8. The color scale is the normalized intensity distribution (amplitude squared) of complex electric field on K1 due to the illumination of K2. The half-intensity contour is indicated with a dashed line.
    }\label{fig:K2_K1_3_1_simplified}
\end{figure}

In Fig.~\ref{fig:K2_K1_3_1_simplified}, we provide an example beam profile in a coordinate frame centered on the middle of the mirror. The method described above is used to generate predicted maps for any combination of wavelength, feedhorn number, or K-mirror position available in the Zemax model. It is worth noting that it is not possible to generate an equivalent profile natively in Zemax, due to the transfer matrix method of propagation that prohibits evaluating the field on the surface of an optic. We then convolve these ``true" beam maps with the emitter pattern spatial profile to arrive at the predicted beam maps.

\clearpage

\section{Camera Calibration} \label{app:cal}

The purpose of camera calibration is to estimate, for a given camera model, the parameters that best describe the images produced by the camera lens projecting a pattern of known geometry onto the sensor. This is essential to accurately remove distortion from images before measuring points of interest in them and reversing the projection operation to infer the 3D location of an object we have imaged. For a full explanation of of camera calibration and projective geometry, we refer the reader to the OpenCV documentation (especially the ``Camera Calibration and 3D Reconstruction" section, whose notation and conventions we follow, though variables may overlap with others in this paper) and Hartley and Zisserman\cite{hartley_multiple_2004}.

We use a pinhole camera model with 8 lens distortion coefficients, 6 radial and 2 tangential. Following OpenCV naming conventions, our calibrated distortion coefficients are
\begin{equation}
\begin{bmatrix}
    k_1\\
    k_2\\
    k_3\\
    k_4\\
    k_5\\
    k_6\\
    p_1\\
    p_2
\end{bmatrix}
    =
\begin{bmatrix}
    -0.00653\\
    -0.144\\
    -0.000366\\
    0.000671\\
    0.0318\\
    -0.00406\\
    -0.143\\
    0.0311
\end{bmatrix},
\end{equation}
$k$ for radial distortion, $p$ for tangential distortion.

To perform calibration, we need to collect information on how our camera projects a known 3D geometry onto the image plane. For our known geometry, we use a ChArUco board: a black and white chessboard pattern interspersed with ArUco coded targets. From images of the calibration board, we estimate the pose relative to the camera, solving the Perspective-n-Point problem. We can then rotate and translate the known chessboard geometry into the estimated pose, project it into the image plane, and compare the chessboard intersections measured from the image against the transformed and reprojected ideal geometry. These errors, known as reprojection errors, are inputs to the Levenberg-Marquardt algorithm to find the best-fit camera parameters in the least-squares sense.

To maximize the constraining power of finite observations, we supply observations that make each parameter evident: we view the calibration board at highly inclined angles to make the effects of focal length obvious, and take many images with the chessboard corners at the edges, where lens distortion is most evident.

\begin{figure}
    \centering
    \includegraphics[width=\textwidth]{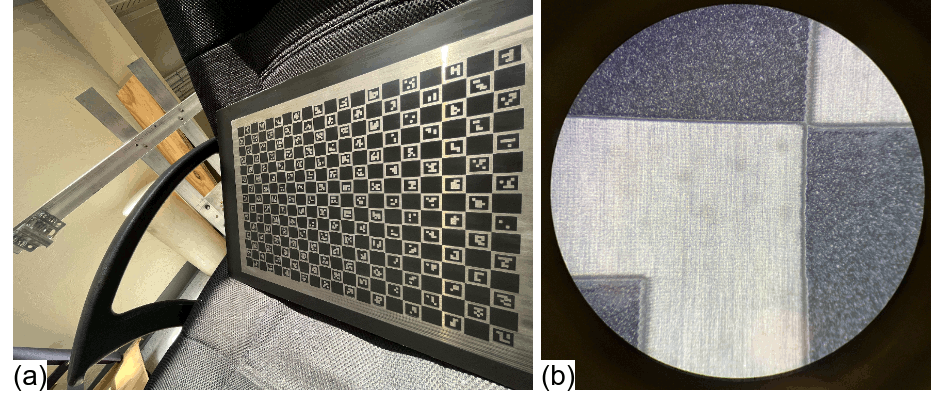}
    \caption{(a) An example image from the calibration dataset, 1 of 46, showing the laser-etched ChArUco board. The board is illuminated from above and angled away from the camera, so that the effect of the wide angle perspective is evident. (b) Microscope image of the laser-etched calibration board region where four squares (two "white" and two black) intersect, forming the one corner that is used in calibration. The laser kerf is evident as slightly fuzzy edges and measures $\sim$10~$\mu$m. The laser scan direction is horizontal in the image, and proceeds in rows from top to bottom.}
    \label{fig:cal_example}
\end{figure}

Our calibration board is laser-etched onto a composite panel of low-density polyethylene (LDPE) clad in black anodized aluminum, which makes a rigid and stable substrate. The CO2 laser etching leaves a satin grey metallic finish, and careful lighting control gives the high contrast necessary for calibration, though a fiber laser would leave a matte white finish. An image from the dataset used to calibrate our camera is shown in Fig.~\ref{fig:cal_example}a, and a microscope image of an intersection is shown in Fig.~\ref{fig:cal_example}b.
The total calibration dataset consists of 46 images with a variety of poses. We take calibration images at a similar distance to the measurement working distance used during the experiment.

\begin{figure}
    \centering
    \includegraphics[width=\textwidth]{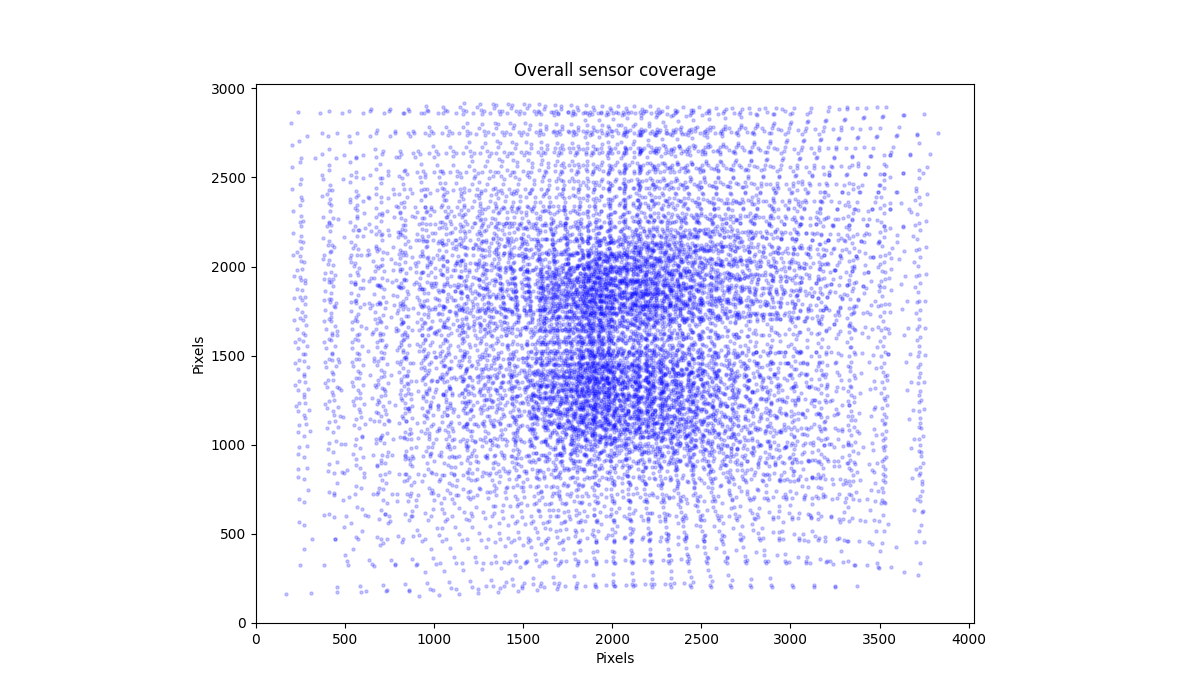}
    \caption{Coverage of identified ChArUco board corners from all calibration images over the image sensor plane. The ideal calibration dataset covers the entire image plane evenly with points, including the edges. Including points at the edges helps to constrain lens distortion parameters.}
    \label{fig:coverage}
\end{figure}

A standard procedure is to check the coverage of calibration data for uniformity over the camera sensor by plotting the image-space position of each identified ChArUco board intersection, which is shown in Fig.~\ref{fig:coverage}.

\begin{figure}
    \centering
    \includegraphics[width=\textwidth]{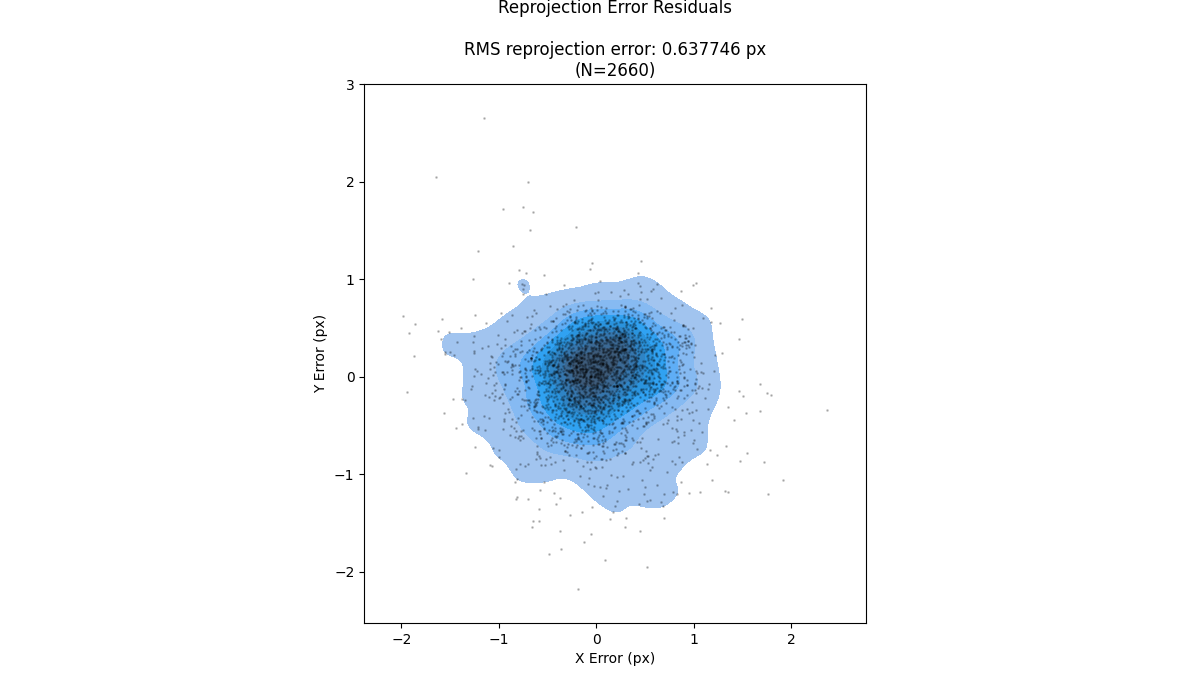}
    \caption{Scatter plot and kernel density estimate of reprojection errors from all calibration images. Each point represents the error vector formed by subtracting a corner's location in the image vs. the ideal ChArUco board corner location in the image plane after applying pose estimation and the estimated camera projection matrix. The ideal set of reprojection errors is circular and Gaussian distributed about the origin, with an RMSE (much) less than a pixel.}
    \label{fig:reproj}
\end{figure}

As another quality check, we show a scatterplot and kernel density estimate of the reprojection errors in Fig.~\ref{fig:reproj}. These are the errors between the de-distorted and reconstructed features as projected onto the sensor vs. the actual measured positions in each image. The errors are roughly Gaussian, circular, and centered on zero, indicating unbiased reprojection errors and good calibration. The root-mean-squared-error across all calibration images, is $\textrm{RPE}_{RMS} = 0.638$~px.

\subsection*{Disclosures}

The authors declare no conflicts of interest.

\subsection*{Acknowledgments}
The authors thank Daewook Kim, Carlos Vargas, Chad Bender, and Erika Hamden for their valuable feedback in the preparation of the manuscript. ECM acknowledges Arash Roshanineshat for helpful discussions in designing the control circuit for the Hawkeye sources, and David Forbes for reviewing the PCB design. ECM also acknowledges the assistance of Dylan Molina, who provided instruction, ideas, and helpful feedback to optimize the laser etched ChArUco board using the Arizona CATalyst Studios. In addition to OpenCV\cite{bradski_opencv_2000}, this research made use of \texttt{numpy}\cite{harris_array_2020}, \texttt{matplotlib}\cite{caswell_matplotlibmatplotlib_2023}, \texttt{pandas}\cite{team_pandas-devpandas_2023}, and \texttt{seaborn}\cite{waskom_mwaskomseaborn_2022}.
The construction of the TIME instrument was supported by the U.S. National Science Foundation ATI award AST-1910598, AAG awards AST-2308041, AST-2308039, AST-2308040, and AST-2308042, and CAREER award AST-1653228.
The UArizona ARO 12-meter Telescope on Kitt Peak is operated by the Arizona Radio Observatory (ARO), Steward Observatory, University of Arizona.

\subsection* {Code, Data, and Materials Availability} 

Stereolithography (STL) files for the 3D printed components, KiCAD and Gerber design files for the printed circuit board (PCB), as well as simulation, control, and measurement post-processing code are all available open source at \linkable{https://github.com/evanmayer/hotspot/}. A tagged version of the control code, analysis notebooks, original calibration images, and original measurement images are available at \linkable{https://doi.org/10.5281/zenodo.10127905}.\cite{mayer_hotspot_2023}


\bibliography{mirror_mapper}   

@article{kino_error_2018,
	title = {Error {Analysis} by {Kinetics} for {Parallel}-{Wire} {Driven} {System} {Using} {Approximated} {Inverse} {Kinematics}},
	volume = {30},
	url = {https://www.fujipress.jp/jrm/rb/robot003000050763/},
	doi = {10.20965/jrm.2018.p0763},
	number = {5},
	urldate = {2021-09-08},
	journal = {Journal of Robotics and Mechatronics},
	author = {Kino, Hitoshi and Imamura, Takumi and Sakagami, Norimitsu},
	month = oct,
	year = {2018},
	note = {Publisher: Fuji Technology Press Ltd.},
	pages = {763--771},
}

@inproceedings{williams_planar_2001,
	address = {Pittsburgh, Pennsylvania, USA},
	title = {Planar {Cable}-{Direct}-{Driven} {Robots}: {Part} {I} — {Kinematics} and {Statics}},
	isbn = {978-0-7918-8023-4},
	shorttitle = {Planar {Cable}-{Direct}-{Driven} {Robots}},
	url = {https://asmedigitalcollection.asme.org/IDETC-CIE/proceedings/IDETC-CIE2001/80234/1233/1090873},
	doi = {10.1115/DETC2001/DAC-21145},
	language = {en},
	urldate = {2021-09-08},
	booktitle = {Volume {2B}: 27th {Design} {Automation} {Conference}},
	publisher = {American Society of Mechanical Engineers},
	author = {Williams, Robert L. and Gallina, Paolo},
	month = sep,
	year = {2001},
	pages = {1233--1240},
}

@article{fast_collaboration_commissioning_2019,
	title = {Commissioning progress of the {FAST}},
	volume = {62},
	issn = {1674-7348, 1869-1927},
	url = {http://link.springer.com/10.1007/s11433-018-9376-1},
	doi = {10.1007/s11433-018-9376-1},
	language = {en},
	number = {5},
	urldate = {2021-11-07},
	journal = {Science China Physics, Mechanics \& Astronomy},
	author = {{FAST Collaboration} and Jiang, Peng and Yue, YouLing and Gan, HengQian and Yao, Rui and Li, Hui and Pan, GaoFeng and Sun, JingHai and Yu, DongJun and Liu, HongFei and Tang, NingYu and Qian, Lei and Lu, JiGuang and Yan, Jun and Peng, Bo and Zhang, ShuXin and Wang, QiMing and Li, Qi and Li, Di},
	month = may,
	year = {2019},
	pages = {959502},
}

@article{kovetz_line-intensity_2017,
	title = {Line-{Intensity} {Mapping}: 2017 {Status} {Report}},
	shorttitle = {Line-{Intensity} {Mapping}},
	url = {http://arxiv.org/abs/1709.09066},
	language = {en},
	urldate = {2021-10-26},
	journal = {arXiv:1709.09066 [astro-ph]},
	author = {Kovetz, Ely D. and Viero, Marco P. and Lidz, Adam and Newburgh, Laura and Rahman, Mubdi and Switzer, Eric and Kamionkowski, Marc and Aguirre, James and Alvarez, Marcelo and Bock, James and Bond, J. Richard and Bower, Goeffry and Bradford, C. Matt and Breysse, Patrick C. and Bull, Philip and Chang, Tzu-Ching and Cheng, Yun-Ting and Chung, Dongwoo and Cleary, Kieran and Corray, Asantha and Crites, Abigail and Croft, Rupert and Doré, Olivier and Eastwood, Michael and Ferrara, Andrea and Fonseca, José and Jacobs, Daniel and Keating, Garrett K. and Lagache, Guilaine and Lakhlani, Gunjan and Liu, Adrian and Moodley, Kavilan and Murray, Norm and Pénin, Aurélie and Popping, Gergö and Pullen, Anthony and Reichers, Dominik and Saito, Shun and Saliwanchik, Ben and Santos, Mario and Somerville, Rachel and Stacey, Gordon and Stein, George and Villaescusa-Navarro, Francesco and Visbal, Eli and Weltman, Amanda and Wolz, Laura and Zemcov, Micheal},
	month = sep,
	year = {2017},
	note = {arXiv: 1709.09066},
}

@inproceedings{marrone_terahertz_2022,
	title = {The terahertz intensity mapper: a balloon-borne imaging spectrometer for galaxy evolution},
	volume = {12190},
	shorttitle = {The terahertz intensity mapper},
	url = {https://www.spiedigitallibrary.org/conference-proceedings-of-spie/12190/1219008/The-terahertz-intensity-mapper--a-balloon-borne-imaging-spectrometer/10.1117/12.2630644.full},
	doi = {10.1117/12.2630644},
	urldate = {2022-10-31},
	booktitle = {Millimeter, {Submillimeter}, and {Far}-{Infrared} {Detectors} and {Instrumentation} for {Astronomy} {XI}},
	publisher = {SPIE},
	author = {Marrone, Daniel P. and Aguirre, James E. and Bracks, Justin S. and Bradford, Charles M. and Brendal, Brockton S. and Bumble, Bruce and Corso, Anthony J. and Devlin, Mark J. and Emerson, Nick and Filippini, Jeffrey P. and Fu, Jianyang and Gasho, Victor and Groppi, Christopher E. and Hailey-Dunsheath, Steve and Hoh, Jonathan and Hollister, Matthew I. and Janssen, Reinier M. J. and Joralmon, Dylan and Keenan, Ryan P. and Liu, Lun-Jun and Lowe, Ian and Mauskopf, Philip and Mayer, Evan C. and Nie, Rong and Razavimaleki, Vesal and Redford, Joseph and Saeid, Talia and Trumper, Isaac L. and Vieira, Joaquin D.},
	month = aug,
	year = {2022},
	pages = {131--142},
}

@article{tong_near_2003,
	title = {Near field vector beam measurements at 1 {THz}},
	volume = {13},
	issn = {1531-1309, 1558-1764},
	url = {http://ieeexplore.ieee.org/document/1208418/},
	doi = {10.1109/LMWC.2003.814602},
	language = {en},
	number = {6},
	urldate = {2022-09-13},
	journal = {IEEE Microwave and Wireless Components Letters},
	author = {Tong, C.-Y.E. and Meledin, D.V. and Marrone, D.P. and Paine, S.N. and Gibson, H. and Blundell, R.},
	month = jun,
	year = {2003},
	pages = {235--237},
}

@article{kim_tilted_2018,
	title = {Tilted {Beam} {Measurement} of {VLBI} {Receiver} for the {South} {Pole} {Telescope}},
	language = {en},
	author = {Kim, Junhan and Marrone, Daniel P},
	year = {2018},
	pages = {5},
}

@inproceedings{miermeister_auto-calibration_2012,
	title = {Auto-{Calibration} {Method} for {Overconstrained} {Cable}-{Driven} {Parallel} {Robots}},
	booktitle = {{ROBOTIK} 2012; 7th {German} {Conference} on {Robotics}},
	author = {Miermeister, Philipp and Pott, Andreas and Verl, Alexander},
	month = may,
	year = {2012},
	pages = {1--6},
}

@inproceedings{jin_four-cable-driven_2013,
	address = {Gwangju, Korea},
	title = {Four-cable-driven parallel robot},
	language = {en},
	author = {Jin, XueJun and Jun, Dae Ik and Pott, Andreas and Park, Sukho and Park, Jong-Oh and Ko, Seong Young},
	year = {2013},
	pages = {6},
}

@article{bernal_line-intensity_2022,
	title = {Line-{Intensity} {Mapping}: {Theory} {Review}},
	volume = {30},
	issn = {0935-4956, 1432-0754},
	shorttitle = {Line-{Intensity} {Mapping}},
	url = {http://arxiv.org/abs/2206.15377},
	doi = {10.1007/s00159-022-00143-0},
	language = {en},
	number = {1},
	urldate = {2023-02-26},
	journal = {The Astronomy and Astrophysics Review},
	author = {Bernal, José Luis and Kovetz, Ely D.},
	month = dec,
	year = {2022},
	note = {arXiv:2206.15377 [astro-ph]},
	pages = {5},
}

@inproceedings{chablat_comparative_2011,
	address = {Washington, DC, USA},
	title = {A {Comparative} {Study} of 4-{Cable} {Planar} {Manipulators} {Based} on {Cylindrical} {Algebraic} {Decomposition}},
	isbn = {978-0-7918-5483-9},
	url = {https://asmedigitalcollection.asme.org/IDETC-CIE/proceedings/IDETC-CIE2011/54839/1253/354326},
	doi = {10.1115/DETC2011-47726},
	language = {en},
	urldate = {2023-02-26},
	booktitle = {Volume 6: 35th {Mechanisms} and {Robotics} {Conference}, {Parts} {A} and {B}},
	publisher = {ASMEDC},
	author = {Chablat, Damien and Ottaviano, Erika and Moroz, Guillaume},
	month = jan,
	year = {2011},
	pages = {1253--1262},
}

@inproceedings{ottaviano_low-cost_2005,
	address = {Barcelona, Spain},
	title = {A {Low}-{Cost} {Easy} {Operation} 4-{Cable} {Driven} {Parallel} {Manipulator}},
	isbn = {978-0-7803-8914-4},
	url = {http://ieeexplore.ieee.org/document/1570734/},
	doi = {10.1109/ROBOT.2005.1570734},
	language = {en},
	urldate = {2023-02-26},
	booktitle = {Proceedings of the 2005 {IEEE} {International} {Conference} on {Robotics} and {Automation}},
	publisher = {IEEE},
	author = {Ottaviano, E. and Ceccarelli, M. and Paone, A. and Carbone, G.},
	year = {2005},
	pages = {4008--4013},
}

@article{jomartov_development_2021,
	title = {Development of a planar cable parallel robot for practical application in the educational process},
	volume = {4},
	issn = {1729-4061, 1729-3774},
	url = {http://journals.uran.ua/eejet/article/view/237772},
	doi = {10.15587/1729-4061.2021.237772},
	language = {en},
	number = {7(112)},
	urldate = {2023-02-26},
	journal = {Eastern-European Journal of Enterprise Technologies},
	author = {Jomartov, Assylbek and Kamal, Aziz and Abduraimov, Azizbek},
	month = aug,
	year = {2021},
	pages = {67--75},
}

@incollection{gosselin_initial_2018,
	address = {Cham},
	title = {Initial {Length} and {Pose} {Calibration} for {Cable}-{Driven} {Parallel} {Robots} with {Relative} {Length} {Feedback}},
	volume = {53},
	isbn = {978-3-319-61430-4 978-3-319-61431-1},
	url = {http://link.springer.com/10.1007/978-3-319-61431-1_13},
	language = {en},
	urldate = {2023-02-26},
	booktitle = {Cable-{Driven} {Parallel} {Robots}},
	publisher = {Springer International Publishing},
	author = {Lau, Darwin},
	editor = {Gosselin, Clément and Cardou, Philippe and Bruckmann, Tobias and Pott, Andreas},
	year = {2018},
	doi = {10.1007/978-3-319-61431-1_13},
	note = {Series Title: Mechanisms and Machine Science},
	pages = {140--151},
}

@inproceedings{wilson_toltec_2020,
	address = {Online Only, United States},
	title = {The {TolTEC} camera: an overview of the instrument and in-lab testing results},
	isbn = {978-1-5106-3693-4 978-1-5106-3694-1},
	shorttitle = {The {TolTEC} camera},
	url = {https://www.spiedigitallibrary.org/conference-proceedings-of-spie/11453/2562331/The-TolTEC-camera--an-overview-of-the-instrument-and/10.1117/12.2562331.full},
	doi = {10.1117/12.2562331},
	language = {en},
	urldate = {2023-02-26},
	booktitle = {Millimeter, {Submillimeter}, and {Far}-{Infrared} {Detectors} and {Instrumentation} for {Astronomy} {X}},
	publisher = {SPIE},
	author = {Wilson, Grant W. and Abi-saad, Sophia and Ade, Peter and Aretxaga, Itziar and Austermann, Jason E. and Ban, Yvonne and Bardin, Joseph and Beall, James A. and Berthoud, Marc and Bryan, Sean A. and Bussan, John and Castillo-Domínguez, Edgar and Chavez, Miguel and Contente, Reid and DeNigris, Natalie W. and Dober, Bradley and Eiben, Miranda and Ferrusca, Daniel and Fissel, Laura and Gao, Jiansong and Golec, Joey and Golina, Robert and Gomez, Arturo and Gordon, Sam and Gutermuth, Robert and Hilton, Gene and Hosseini, Mohsen and Hubmayr, Johannes and Hughes, David and Kuczarski, Stephen W. and Lee, Dennis and Lunde, Emily and Ma, Zhiyuan and Mani, Hamdi and Mauskopf, Philip and McCrackan, Michael and McKenney, Christopher and McMahon, Jeffrey and Novak, Giles and Pisano, Giampaolo and Pope, Alexandra and Ralston, Amy and Rodriguez, Ivan and Sánchez-Argüelles, David and Schloerb, F. Peter and Simon, Sara M. and Sinclair, Adrian and Souccar, Kamal and Torres Campos, Ana and Tucker, Carole and Ullom, Joel and Van Camp, Eric and Van Lanen, Jeff and Velazquez, Miguel and Vissers, Michael and Weeks, Eric and Yun, Min S.},
	editor = {Zmuidzinas, Jonas and Gao, Jian-Rong},
	month = dec,
	year = {2020},
	pages = {1},
}

@article{borgstrom_nims-pl_2009,
	title = {{NIMS}-{PL}: {A} {Cable}-{Driven} {Robot} {With} {Self}-{Calibration} {Capabilities}},
	volume = {25},
	issn = {1552-3098, 1941-0468},
	shorttitle = {{NIMS}-{PL}},
	url = {http://ieeexplore.ieee.org/document/5159377/},
	doi = {10.1109/TRO.2009.2024792},
	language = {en},
	number = {5},
	urldate = {2023-02-26},
	journal = {IEEE Transactions on Robotics},
	author = {Borgstrom, P.H. and Jordan, B.L. and Borgstrom, B.J. and Stealey, M.J. and Sukhatme, G.S. and Batalin, M.A. and Kaiser, W.J.},
	month = oct,
	year = {2009},
	pages = {1005--1015},
}

@article{williams_planar_2003,
	title = {Planar {Translational} {Cable}-{Direct}-{Driven} {Robots}},
	volume = {20},
	issn = {0741-2223, 1097-4563},
	url = {https://onlinelibrary.wiley.com/doi/10.1002/rob.10073},
	doi = {10.1002/rob.10073},
	language = {en},
	number = {3},
	urldate = {2023-02-26},
	journal = {Journal of Robotic Systems},
	author = {Williams, Robert L. and Gallina, Paolo and Vadia, Jigar},
	month = mar,
	year = {2003},
	pages = {107--120},
}

@patent{brown_suspension_1987,
	title = {Suspension system for supporting and conveying equipment, such as a camera},
	url = {https://patents.google.com/patent/US4710819A/en},
	nationality = {US},
	assignee = {Brown Garrett W},
	number = {US4710819A},
	urldate = {2023-06-07},
	author = {Brown, Garrett W.},
	month = dec,
	year = {1987},
}

@phdthesis{landsberger_design_1984,
	type = {Thesis},
	title = {Design and construction of a cable-controlled, parallel link manipulator},
	copyright = {M.I.T. theses are protected by copyright. They may be viewed from this source for any purpose, but reproduction or distribution in any format is prohibited without written permission. See provided URL for inquiries about permission.},
	url = {https://dspace.mit.edu/handle/1721.1/15333},
	language = {eng},
	urldate = {2023-06-07},
	school = {Massachusetts Institute of Technology},
	author = {Landsberger, Samuel Ernest},
	year = {1984},
	note = {Accepted: 2005-08-05T19:30:01Z},
}

@article{stewart_platform_1965,
	title = {A {Platform} with {Six} {Degrees} of {Freedom}},
	volume = {180},
	issn = {0020-3483},
	url = {https://doi.org/10.1243/PIME_PROC_1965_180_029_02},
	doi = {10.1243/PIME_PROC_1965_180_029_02},
	language = {en},
	number = {1},
	urldate = {2023-06-07},
	journal = {Proceedings of the Institution of Mechanical Engineers},
	author = {Stewart, D.},
	month = jun,
	year = {1965},
	note = {Publisher: IMECHE},
	pages = {371--386},
}

@patent{cappel_motion_1967,
	title = {Motion simulator},
	url = {https://patents.google.com/patent/US3295224A/en},
	nationality = {US},
	assignee = {Franklin Institute},
	number = {US3295224A},
	urldate = {2023-06-07},
	author = {Cappel, Klaus L.},
	month = jan,
	year = {1967},
}

@book{gough_universal_1962,
	title = {Universal tyre test machine},
	volume = {Institution of Mechanical Engineers},
	language = {eng},
	author = {Gough, V. E. and Whitehall, S. G.},
	year = {1962},
	note = {Pages: p. 117-137
Publication Title: Eley, G., ed., FISITA. International Automobile Technical Congress. Ninth. Proceedings
OCLC: 173370995},
}

@inproceedings{lowe_characterization_2020,
	title = {Characterization, deployment, and in-flight performance of the {BLAST}-{TNG} cryogenic receiver},
	url = {http://arxiv.org/abs/2012.01372},
	doi = {10.1117/12.2560854},
	language = {en},
	urldate = {2023-06-26},
	booktitle = {Millimeter, {Submillimeter}, and {Far}-{Infrared} {Detectors} and {Instrumentation} for {Astronomy} {X}},
	author = {Lowe, Ian and Ade, Peter A. R. and Ashton, Peter C. and Austermann, Jason E. and Coppi, Gabriele and Cox, Erin G. and Devlin, Mark J. and Dober, Bradley J. and Fanfani, Valentina and Fissel, Laura M. and Galitzki, Nicholas and Gao, Jiansong and Gordon, Samuel and Groppi, Christopher E. and Hilton, Gene C. and Hubmayr, Johannes and Klein, Jeffrey and Li, Dale and Lourie, Nathan P. and Mani, Hamdi and Mauskopf, Philip and McKenney, Christopher and Nati, Federico and Novak, Giles and Pisano, Giampaolo and Romualdez, L. Javier and Soler, Juan D. and Sinclair, Adrian and Tucker, Carole and Ullom, Joel and Vissers, Michael and Wheeler, Caleb and Williams, Paul A.},
	month = dec,
	year = {2020},
	note = {arXiv:2012.01372 [astro-ph]},
	pages = {3},
}

@article{trumper_freeform_2019,
	title = {Freeform surface selection based on parametric fitness function using modal wavefront fitting},
	volume = {27},
	issn = {1094-4087},
	url = {https://opg.optica.org/abstract.cfm?URI=oe-27-5-6815},
	doi = {10.1364/OE.27.006815},
	language = {en},
	number = {5},
	urldate = {2023-06-26},
	journal = {Optics Express},
	author = {Trumper, Isaac and Aftab, Maham and Kim, Dae Wook},
	month = mar,
	year = {2019},
	pages = {6815},
}

@inproceedings{hunacek_detector_2016,
	title = {Detector modules and spectrometers for the {TIME}-{Pilot} [{CII}] intensity mapping experiment},
	volume = {9914},
	url = {https://www.spiedigitallibrary.org/conference-proceedings-of-spie/9914/99140L/Detector-modules-and-spectrometers-for-the-TIME-Pilot-CII-intensity/10.1117/12.2233762.full},
	doi = {10.1117/12.2233762},
	urldate = {2023-06-27},
	booktitle = {Millimeter, {Submillimeter}, and {Far}-{Infrared} {Detectors} and {Instrumentation} for {Astronomy} {VIII}},
	publisher = {SPIE},
	author = {Hunacek, Jonathon and Bock, James and Bradford, C. Matt and Bumble, Bruce and Chang, Tzu-Ching and Cheng, Yun-Ting and Cooray, Asantha and Crites, Abigail and Hailey-Dunsheath, Steven and Gong, Yan and Li, Chao-Te and O’Brient, Roger and Shirokoff, Erik and Shiu, Corwin and Sun, Jason and Staniszewski, Zachary and Uzgil, Bade and Zemcov, Michael},
	month = jul,
	year = {2016},
	pages = {132--141},
}

@inproceedings{li_time_2018,
	address = {Austin, United States},
	title = {{TIME} millimeter wave grating spectrometer},
	isbn = {978-1-5106-1969-2 978-1-5106-1970-8},
	url = {https://spiedigitallibrary.org/conference-proceedings-of-spie/10708/2311415/TIME-millimeter-wave-grating-spectrometer/10.1117/12.2311415.full},
	doi = {10.1117/12.2311415},
	language = {en},
	urldate = {2023-06-27},
	booktitle = {Millimeter, {Submillimeter}, and {Far}-{Infrared} {Detectors} and {Instrumentation} for {Astronomy} {IX}},
	publisher = {SPIE},
	author = {Li, Chao-Te and Chang, Tzu-Ching and Bradford, Charles M. and Crites, Abigail and Hunacek, Jonathon and Bock, James J. and Wei, Ta-Shun and Cheng, Jen-Chieh},
	editor = {Zmuidzinas, Jonas and Gao, Jian-Rong},
	month = jul,
	year = {2018},
	pages = {114},
}

@article{garrido-jurado_automatic_2014,
	title = {Automatic generation and detection of highly reliable fiducial markers under occlusion},
	volume = {47},
	issn = {0031-3203},
	url = {https://www.sciencedirect.com/science/article/pii/S0031320314000235},
	doi = {10.1016/j.patcog.2014.01.005},
	language = {en},
	number = {6},
	urldate = {2023-08-08},
	journal = {Pattern Recognition},
	author = {Garrido-Jurado, S. and Muñoz-Salinas, R. and Madrid-Cuevas, F. J. and Marín-Jiménez, M. J.},
	month = jun,
	year = {2014},
	pages = {2280--2292},
}

@article{bradski_opencv_2000,
	title = {The {OpenCV} {Library}},
	journal = {Dr. Dobb's Journal of Software Tools},
	author = {Bradski, G.},
	year = {2000},
}

@article{zhang_flexible_2000,
	title = {A flexible new technique for camera calibration},
	volume = {22},
	issn = {1939-3539},
	doi = {10.1109/34.888718},
	number = {11},
	journal = {IEEE Transactions on Pattern Analysis and Machine Intelligence},
	author = {Zhang, Z.},
	month = nov,
	year = {2000},
	note = {Conference Name: IEEE Transactions on Pattern Analysis and Machine Intelligence},
	pages = {1330--1334},
}

@article{marchand_pose_2016,
	title = {Pose {Estimation} for {Augmented} {Reality}: {A} {Hands}-{On} {Survey}},
	volume = {22},
	shorttitle = {Pose {Estimation} for {Augmented} {Reality}},
	url = {https://inria.hal.science/hal-01246370},
	doi = {10.1109/TVCG.2015.2513408},
	language = {en},
	number = {12},
	urldate = {2023-08-08},
	journal = {IEEE Transactions on Visualization and Computer Graphics},
	author = {Marchand, Eric and Uchiyama, Hideaki and Spindler, Fabien},
	year = {2016},
	pages = {2633},
}

@inproceedings{wang_apriltag_2016,
	title = {{AprilTag} 2: {Efficient} and robust fiducial detection},
	shorttitle = {{AprilTag} 2},
	doi = {10.1109/IROS.2016.7759617},
	booktitle = {2016 {IEEE}/{RSJ} {International} {Conference} on {Intelligent} {Robots} and {Systems} ({IROS})},
	author = {Wang, John and Olson, Edwin},
	month = oct,
	year = {2016},
	note = {ISSN: 2153-0866},
	pages = {4193--4198},
}

@inproceedings{pentenrieder_analysis_2007,
	title = {Analysis of {Tracking} {Accuracy} for {Single}-{Camera} {Square}-{Marker}-{Based} {Tracking}},
	url = {https://www.semanticscholar.org/paper/Analysis-of-Tracking-Accuracy-for-Single-Camera-Pentenrieder/70c5d9b33a978ff2d03eeaa627afaf4f6f609a1f},
	urldate = {2023-08-15},
	author = {Pentenrieder, Katharina},
	year = {2007},
}

@phdthesis{hunacek_time_2018,
	title = {{TIME}: {A} {Millimeter}-{Wavelength} {Grating} {Spectrometer} {Array} for [{CII}] / {CO} {Intensity} {Mapping}},
	url = {http://link.springer.com/10.1007/s10909-018-1906-3},
	language = {en},
	urldate = {2023-10-14},
	author = {Hunacek, J.},
	month = dec,
	year = {2018},
}

@article{sun_probing_2021,
	title = {Probing {Cosmic} {Reionization} and {Molecular} {Gas} {Growth} with {TIME}},
	volume = {915},
	issn = {0004-637X, 1538-4357},
	url = {http://arxiv.org/abs/2012.09160},
	doi = {10.3847/1538-4357/abfe62},
	language = {en},
	number = {1},
	urldate = {2023-10-17},
	journal = {The Astrophysical Journal},
	author = {Sun, Guochao and Chang, Tzu-Ching and Uzgil, Bade D. and Bock, Jamie and Bradford, Charles M. and Butler, Victoria and Caze-Cortes, Tessalie and Cheng, Yun-Ting and Cooray, Asantha and Crites, Abigail T. and Hailey-Dunsheath, Steve and Emerson, Nick and Frez, Clifford and Hoscheit, Benjamin L. and Hunacek, Jonathon R. and Keenan, Ryan P. and Li, Chao-Te and Madonia, Paolo and Marrone, Daniel P. and Moncelsi, Lorenzo and Shiu, Corwin and Trumper, Isaac and Turner, Anthony and Weber, Alexis and Wei, Ta-Shun and Zemcov, Michael},
	month = jul,
	year = {2021},
	note = {arXiv:2012.09160 [astro-ph]},
	pages = {33},
}

@inproceedings{crites_time-pilot_2014,
	title = {The {TIME}-{Pilot} intensity mapping experiment},
	volume = {9153},
	url = {https://www.spiedigitallibrary.org/conference-proceedings-of-spie/9153/91531W/The-TIME-Pilot-intensity-mapping-experiment/10.1117/12.2057207.full},
	doi = {10.1117/12.2057207},
	urldate = {2023-11-06},
	booktitle = {Millimeter, {Submillimeter}, and {Far}-{Infrared} {Detectors} and {Instrumentation} for {Astronomy} {VII}},
	publisher = {SPIE},
	author = {Crites, A. T. and Bock, J. J. and Bradford, C. M. and Chang, T. C. and Cooray, A. R. and Duband, L. and Gong, Y. and Hailey-Dunsheath, S. and Hunacek, J. and Koch, P. M. and Li, C. T. and O'Brient, R. C. and Prouve, T. and Shirokoff, E. and Silva, M. B. and Staniszewski, Z. and Uzgil, B. and Zemcov, M.},
	month = aug,
	year = {2014},
	pages = {613--621},
}

@article{hunacek_design_2016,
	title = {Design and {Fabrication} of {TES} {Detector} {Modules} for the {TIME}-{Pilot} [{CII}] {Intensity} {Mapping} {Experiment}},
	volume = {184},
	issn = {1573-7357},
	url = {https://doi.org/10.1007/s10909-015-1359-x},
	doi = {10.1007/s10909-015-1359-x},
	language = {en},
	number = {3},
	urldate = {2023-11-06},
	journal = {Journal of Low Temperature Physics},
	author = {Hunacek, J. and Bock, J. and Bradford, C. M. and Bumble, B. and Chang, T.-C. and Cheng, Y.-T. and Cooray, A. and Crites, A. and Hailey-Dunsheath, S. and Gong, Y. and Kenyon, M. and Koch, P. and Li, C.-T. and O’Brient, R. and Shirokoff, E. and Shiu, C. and Staniszewski, Z. and Uzgil, B. and Zemcov, M.},
	month = aug,
	year = {2016},
	pages = {733--738},
}

@article{albus_nist_1992,
	title = {The {NIST} {SPIDER}, {A} {Robot} {Crane}},
	volume = {97},
	issn = {1044-677X},
	doi = {10.6028/jres.097.016},
	language = {eng},
	number = {3},
	journal = {Journal of Research of the National Institute of Standards and Technology},
	author = {Albus, James and Bostelman, Roger and Dagalakis, Nicholas},
	year = {1992},
	pmid = {28053439},
	pmcid = {PMC4909171},
	pages = {373--385},
}

@article{perrin_poppy_2016,
	title = {{POPPY}: {Physical} {Optics} {Propagation} in {PYthon}},
	shorttitle = {{POPPY}},
	url = {https://ui.adsabs.harvard.edu/abs/2016ascl.soft02018P},
	urldate = {2023-11-06},
	journal = {Astrophysics Source Code Library},
	author = {Perrin, Marshall and Long, Joseph and Douglas, Ewan and Sivaramakrishnan, Anand and Slocum, Christine and {others}},
	month = feb,
	year = {2016},
	note = {ADS Bibcode: 2016ascl.soft02018P},
	pages = {ascl:1602.018},
}

@book{hartley_multiple_2004,
	edition = {Second},
	title = {Multiple {View} {Geometry} in {Computer} {Vision}},
	publisher = {Cambridge University Press, ISBN: 0521540518},
	author = {Hartley, R. I. and Zisserman, A.},
	year = {2004},
}

@article{siringo_large_2009,
	title = {The {Large} {APEX} {Bolometer} {Camera} {LABOCA}},
	volume = {497},
	doi = {10.1051/0004-6361/200811454},
	journal = {Astronomy and Astrophysics},
	author = {Siringo, Giorgio and Kreysa, Ernst and Kovacs, Andras and Schuller, Frederic and Weiß, Axel and Esch, Walter and Gemuend, H. and Jethava, Nileshg and Lundershausen, Gundula and Colin, Angel and Güsten, Rolf and Menten, Karl and Beelen, Alexandre and Bertoldi, Frank and Beeman, Jeffrey and Haller, E.},
	month = mar,
	year = {2009},
	pages = {945--962},
}

@inproceedings{gom_testing_2010,
	title = {Testing results and current status of {FTS}-2, an imaging {Fourier} transform spectrometer for {SCUBA}-2},
	volume = {7741},
	url = {https://www.spiedigitallibrary.org/conference-proceedings-of-spie/7741/77412E/Testing-results-and-current-status-of-FTS-2-an-imaging/10.1117/12.857667.full},
	doi = {10.1117/12.857667},
	urldate = {2023-11-09},
	booktitle = {Millimeter, {Submillimeter}, and {Far}-{Infrared} {Detectors} and {Instrumentation} for {Astronomy} {V}},
	publisher = {SPIE},
	author = {Gom, Brad and Naylor, David},
	month = jul,
	year = {2010},
	pages = {710--721},
}

@article{davis_complex_2019,
	title = {Complex {Field} {Mapping} of {Large} {Direct} {Detector} {Focal} {Plane} {Arrays}},
	volume = {9},
	issn = {2156-3446},
	url = {https://ieeexplore.ieee.org/document/8550802},
	doi = {10.1109/TTHZ.2018.2883820},
	number = {1},
	urldate = {2023-11-10},
	journal = {IEEE Transactions on Terahertz Science and Technology},
	author = {Davis, Kristina K. and Yates, Stephen J. C. and Jellema, Willem and Groppi, Christopher E. and Baselmans, Jochem J. A. and Kohno, Kotaro and Baryshev, Andrey M.},
	month = jan,
	year = {2019},
	note = {Conference Name: IEEE Transactions on Terahertz Science and Technology},
	pages = {67--77},
}

@misc{team_pandas-devpandas_2023,
	title = {pandas-dev/pandas: {Pandas}},
	url = {https://doi.org/10.5281/zenodo.10107975},
	publisher = {Zenodo},
	author = {team, The pandas development},
	month = nov,
	year = {2023},
	doi = {10.5281/zenodo.10107975},
}

@misc{caswell_matplotlibmatplotlib_2023,
	title = {matplotlib/matplotlib: {REL}: v3.7.1},
	url = {https://doi.org/10.5281/zenodo.7697899},
	publisher = {Zenodo},
	author = {Caswell, Thomas A. and Lee, Antony and Andrade, Elliott Sales de and Droettboom, Michael and Hoffmann, Tim and Klymak, Jody and Hunter, John and Firing, Eric and Stansby, David and Varoquaux, Nelle and Nielsen, Jens Hedegaard and Root, Benjamin and May, Ryan and Gustafsson, Oscar and Elson, Phil and Seppänen, Jouni K. and Lee, Jae-Joon and Dale, Darren and {hannah} and McDougall, Damon and Straw, Andrew and Hobson, Paul and Sunden, Kyle and Lucas, Greg and Gohlke, Christoph and Vincent, Adrien F. and Yu, Tony S. and Ma, Eric and Silvester, Steven and Moad, Charlie},
	month = mar,
	year = {2023},
	doi = {10.5281/zenodo.7697899},
}

@misc{waskom_mwaskomseaborn_2022,
	title = {mwaskom/seaborn: v0.12.2 ({December} 2022)},
	url = {https://doi.org/10.5281/zenodo.7495530},
	publisher = {Zenodo},
	author = {Waskom, Michael and Gelbart, Maoz and Botvinnik, Olga and Ostblom, Joel and Hobson, Paul and Lukauskas, Saulius and Gemperline, David C. and Augspurger, Tom and Halchenko, Yaroslav and Warmenhoven, Jordi and Cole, John B. and Hoeven, Ewout ter and Ruiter, Julian de and Vanderplas, Jake and Hoyer, Stephan and Pye, Cameron and Miles, Alistair and Swain, Corban and Meyer, Kyle and Martin, Marcel and Bachant, Pete and Molin, Stefanie and Quintero, Eric and Kunter, Gero and Villalba, Santi and {Brian} and Fitzgerald, Clark and Evans, Constantine and Williams, Mike Lee and O'Kane, Drew},
	month = dec,
	year = {2022},
	doi = {10.5281/zenodo.7495530},
}

@article{harris_array_2020,
	title = {Array programming with {NumPy}},
	volume = {585},
	url = {https://doi.org/10.1038/s41586-020-2649-2},
	doi = {10.1038/s41586-020-2649-2},
	number = {7825},
	journal = {Nature},
	author = {Harris, Charles R. and Millman, K. Jarrod and Walt, Stéfan J. van der and Gommers, Ralf and Virtanen, Pauli and Cournapeau, David and Wieser, Eric and Taylor, Julian and Berg, Sebastian and Smith, Nathaniel J. and Kern, Robert and Picus, Matti and Hoyer, Stephan and Kerkwijk, Marten H. van and Brett, Matthew and Haldane, Allan and Río, Jaime Fernández del and Wiebe, Mark and Peterson, Pearu and Gérard-Marchant, Pierre and Sheppard, Kevin and Reddy, Tyler and Weckesser, Warren and Abbasi, Hameer and Gohlke, Christoph and Oliphant, Travis E.},
	month = sep,
	year = {2020},
	note = {Publisher: Springer Science and Business Media LLC},
	pages = {357--362},
}

@misc{mayer_hotspot_2023,
	title = {hotspot: a planar cable-driven parallel robot for submillimeter and terahertz beam mapping measurements},
	url = {https://doi.org/10.5281/zenodo.10127905},
	publisher = {Zenodo},
	author = {Mayer, Evan},
	month = nov,
	year = {2023},
	doi = {10.5281/zenodo.10127905},
}

@article{trumper_utilizing_2019,
	title = {Utilizing freeform optics in dynamic optical configuration designs},
	volume = {5},
	issn = {2329-4124, 2329-4221},
	url = {https://www.spiedigitallibrary.org/journals/Journal-of-Astronomical-Telescopes-Instruments-and-Systems/volume-5/issue-3/035005/Utilizing-freeform-optics-in-dynamic-optical-configuration-designs/10.1117/1.JATIS.5.3.035005.full},
	doi = {10.1117/1.JATIS.5.3.035005},
	number = {3},
	urldate = {2023-12-04},
	journal = {Journal of Astronomical Telescopes, Instruments, and Systems},
	author = {Trumper, Isaac L. and Marrone, Daniel P. and Kim, Dae Wook},
	month = jul,
	year = {2019},
	note = {Publisher: SPIE},
	pages = {035005},
}

@article{fischler_random_1981,
	title = {Random sample consensus: a paradigm for model fitting with applications to image analysis and automated cartography},
	volume = {24},
	issn = {0001-0782},
	shorttitle = {Random sample consensus},
	url = {https://dl.acm.org/doi/10.1145/358669.358692},
	doi = {10.1145/358669.358692},
	number = {6},
	urldate = {2025-07-09},
	journal = {Commun. ACM},
	author = {Fischler, Martin A. and Bolles, Robert C.},
	month = jun,
	year = {1981},
	pages = {381--395},
}
\bibliographystyle{spiejour}   





\listoffigures
\listoftables

\end{document}